\documentclass[preprint]{aastex61}
\usepackage{natbib}
\usepackage{color}

\newcommand{\Msol}{M$_{\odot}$}

\newcommand{\mum}{$\mu m$}

\defcitealias{Dunham16}{D16}

\begin{document}
\title{ALMA Observations of Starless Core Substructure in Ophiuchus}
\author{H. Kirk}
\affil{NRC Herzberg Astronomy and Astrophysics, 5071 West Saanich Rd, Victoria, BC, V9E 2E7, Canada} 

\author{M.~M. Dunham}
\affil{Department of Physics, State University of New York at Fredonia, 280 Central Ave, Fredonia, NY 14063, USA}
\affil{Harvard-Smithsonian Center for Astrophysics, 60 Garden Street, MS 78, Cambridge, MA 02138, USA}

\author{J. Di~Francesco}
\affil{NRC Herzberg Astronomy and Astrophysics, 5071 West Saanich Rd, Victoria, BC, V9E 2E7, Canada} 
\affil{Department of Physics and Astronomy, University of Victoria, Victoria, BC, V8P 1A1, Canada}

\author{D. Johnstone}
\affil{NRC Herzberg Astronomy and Astrophysics, 5071 West Saanich Rd, Victoria, BC, V9E 2E7, Canada}
\affil{Department of Physics and Astronomy, University of Victoria, Victoria, BC, V8P 1A1, Canada}

\author{S.~S.~R. Offner}
\affil{Department of Astronomy, University of Massachusetts, Amherst, MA 01002, USA}

\author{S.~I. Sadavoy}
\affil{Harvard-Smithsonian Center for Astrophysics, 60 Garden Street, MS 78, Cambridge, MA 02138, USA}

\author{J.~J. Tobin}
\affil{Department of Physics and Astronomy, University of Oklahoma, 440 W. Brooks St., 231 Nielsen Hall, Norman, OK 73019, USA}

\author{H.~G. Arce}
\affil{Department of Astronomy, Yale University, P.O. Box 208101, New Haven, CT 06520, USA 0000-0001-5653-7817}

\author{T.~L. Bourke}
\affil{SKA Organization, Jodrell Bank Observatory, Lower Withington, Macclesfield, Cheshire SK11 9DL, UK}

\author{S. Mairs}
\affil{Department of Physics and Astronomy, University of Victoria, Victoria, BC, V8P 1A1, Canada}

\author{P.~C. Myers}
\affil{Harvard-Smithsonian Center for Astrophysics, 60 Garden Street, MS 78, Cambridge, MA 02138, USA}

\author{J.~E. Pineda}
\affil{Max-Planck-Institut fÃ¼r extraterrestrische Physik, Giessenbachstrasse 1, D-85748, Garching, Germany}

\author{S. Schnee}
\affil{National Radio Astronomy Observatory, 520 Edgemont Road, Charlottesville, VA 22903, USA}

\author{Y.~L. Shirley}
\affil{Steward Observatory, University of Arizona, 933 North Cherry Avenue, Tucson, AZ 85721, USA}

\begin{abstract}
Compact substructure is expected to arise in a starless core
as mass becomes concentrated in the central region likely to form a protostar.  
Additionally, multiple peaks 
may form if 
fragmentation occurs.
We present ALMA Cycle 2 observations of 60 starless and protostellar
cores in the Ophiuchus molecular cloud.  
We detect eight compact substructures which are $>15$\arcsec\ 
from the nearest {\it Spitzer} YSO.  Only one of these 
has strong evidence for
being truly starless after considering ancillary data, e.g., from
{\it Herschel} and X-ray telescopes.  An additional 
extended emission structure has tentative evidence for starlessness.
The number of our detections is consistent
with estimates from a combination of synthetic observations
of numerical simulations and analytical arguments.  
This result suggests that a similar
ALMA study in the Chamaeleon~I cloud, which detected no compact 
substructure in starless cores, may
be due to the peculiar evolutionary state of cores in that cloud.
\end{abstract}

\section{Introduction}
Protostars form within dense cores, which in turn are usually found embedded
within larger molecular cloud complexes.  
Furthermore, many stars are found in binary or higher-order multiple systems 
\citep[roughly half of solar-type stars, e.g.,][]{DuquennoyMayor91, Raghavan10, Duchene13}, 
and there is strong evidence to suggest that
multiplicity may be more common in younger protostellar systems 
\citep[e.g.,][]{Looney00,Connelley08b,Chen13,Tobin16a}.
A high protostellar multiplicity is also expected based on models comparing the core mass
function (CMF) and initial stellar mass function (IMF) \citep{Goodwin08,Holman13,Lomax15}.
While the formation of binary and higher order protostellar systems could
arise through fragmentation of a protostellar disk \citep[e.g.,][]{Kratter16,Tobin16c},
some of these multiple systems could also begin forming earlier, at the starless 
core stage.  If so, then dense cores likely have a more complex structure than
the traditional picture of a radially symmetric, smooth density profile which is flat in 
the centre and decreases as a roughly $r^{-2}$ power law on the outskirts 
\citep[e.g.,][]{DiFrancesco07}.  

One mechanism for generating anisotropic structure in cores is turbulence.  In
the model of turbulent fragmentation, for example, turbulence creates denser sub-regions 
within a starless core.  These denser sub-regions can
then individually become Jeans unstable and collapse, later forming multiple
systems with larger (1000's of au) separations \citep{Fisher04,Goodwin04,Offner10,Offner16}.
Interferometers, especially the Atacama Large Millimeter/submillimeter Array (ALMA), 
should be sensitive to substructures generated by processes such as turbulent 
fragmentation for at least the later stages of a starless core's evolution, just before the 
protostar forms \citep{Offner12,Mairs14,Dunham16}.

The detection rate of such compact substructures in starless cores yields
clues to the processes driving their formation.
For example, \citet{Dunham16} find that for an ALMA observing set-up mimicking their
observations of starless cores in Chamaeleon~I (discussed in more detail below),
compact structures generated in starless cores by turbulence should have
a detection rate more than 10 times higher than comparable detections of simple
smooth Bonnor Ebert sphere-like \citep{Bonnor56,Ebert55} cores.  
This difference is attributable to the much smoother density profile of the Bonnor Ebert sphere
than the structures generated in the numerical simulations.
It is only at very high central densities that there is sufficient variation in the small 
scale density structure of the Bonnor Ebert sphere to be visible after interferometry 
filters out the larger-scale emission structures.

The search for compact substructure within starless cores with interferometers
has typically targeted sources with signs of being near protostellar collapse
\citep[e.g.,][]{JKirk09,ChenArce10,Pineda11b,Pineda15,Bourke12,Nakamura12,Lee13,Friesen14,
Kainulainen16}.
Many of these studies focussed on isolated Bok globules rather than on starless
cores within larger-scale clouds, and all but two of those have small samples
of one or two sources.
The two exceptions are \citet{Lee13} who surveyed eight cores but only detected
compact substructure in molecular line emission and not continuum emission, and 
\citet{Kainulainen16} discussed in more detail in Section~5.2.
Larger continuum surveys of many starless cores in the same cloud tend to 
yield a low detection rate.  \citet{Schnee10} initially reported two 
detections in a sample
of 11 starless cores in Perseus, but both of these objects were later shown to
be very low luminosity protostars \citep{Schnee12a}.
These non-detections imply that most of the starless cores surveyed have smooth, extended
structure which would be filtered out by interferometric observations.

Before the advent of ALMA and the Jansky Very Large Array (JVLA), 
interferometric observations of dense cores required a significant
investment of telescope time per core, therefore limiting the above-mentioned studies to
relatively small numbers of targets, with limited sensitivity to substructure on a range
of scales.  Using synthetic observations of numerical simulations of star formation under
the turbulent fragmentation scenario, \citet{Offner12} demonstrated that 
ALMA should have the sensitivity to detect core substructure, especially
when the full array is available and the observations include data from the ALMA 
Compact Array that trace slightly larger angular scales.  The JVLA is similarly 
being used as an effective probe of protostellar multiplicity \citep{Tobin16a}.
ALMA and JVLA observations additionally require significantly less integration time to 
reach the same sensitivity as earlier interferometric arrays, making larger population
studies more easily achievable.

To date, the only ALMA survey of starless cores was published 
by \citet{Dunham16}, hereafter \citetalias{Dunham16}.  \citetalias{Dunham16}
presented ALMA Cycle 1 observations of the entire population of dense cores 
in the Chamaeleon~I cloud, 73 sources in total, at a distance of 150~pc, 
of which none of the 56 starless cores
were detected.  Based on simulations of turbulent cores, discussed in 
more detail in Section~5,
\citetalias{Dunham16} expected at least two detections of compact substructure in
starless cores. 
To reconcile the
lack of detections with the expected number, \citetalias{Dunham16} 
argued that Chamaeleon~I may not be continuously forming stars.  If all of the starless cores
in Chamaeleon~I are at an early evolutionary phase or will dissipate rather than
collapse to form protostars, then it is unlikely any substructure
is sufficiently well-developed to be observable there.
Alternatively, if the starless cores in Chamaeleon~I can be well-described by a 
simple, smooth Bonnor Ebert sphere model, no detections would be expected.
Observations of starless core populations in additional molecular clouds are therefore
crucial to determine whether a `typical' population of dense cores 
have structures that are more complex than the simple Bonnor Ebert sphere model.

The Ophiuchus molecular cloud presents an optimal environment to extend the 
\citetalias{Dunham16}
study.  Ophiuchus is one of the nearest molecular clouds, at only 
$\sim$140~pc \citep{OrtizLeon17}, similar to Chamaeleon.
Its dense core population has been characterized in the (sub)millimetre
continuum by a variety of authors
\citep{Motte98,Johnstone00b,Johnstone04,Stanke06,Young06,Nutter06,Jorgensen08,Enoch08,Pattle15}, 
and its YSO population has been characterized
with {\it Spitzer} \citep{Padgett08,Jorgensen08,Dunham08,Sadavoy10a}.

Here, we present ALMA Cycle 2 observations of the dense cores in Ophiuchus to search
for signs of compact substructure.  
We note that the angular resolution of our observations would allow us to 
see multiple peaks generated under a turbulent fragmentation type of model (at 1000's of
au scales), in addition to compact peaks, which still allow us to test for the
action of processes beyond the simple evolution of a Bonnor Ebert sphere.
Our resolution is insufficient, however, to reveal small separation
multiples (at 100's of au), which might form via disk fragmentation.

We describe our observations and existing applicable data in Section~2, and present
the ALMA detections and their basic properties in Section~3.
In Section~4, we compare the properties of candidate starless detections with
their protostellar counterparts.
In Section~5, we interpret the candidate starless core detections in the context of the 
turbulent fragmentation model, and in Section~6, we summarize our conclusions.

\section{Observations}
\label{sec_obs}

\subsection{Target Selection: SCUBA}
In their joint analysis of SCUBA 850~\mum\ and multi-wavelength {\it Spitzer} data, 
\citet{Jorgensen08} identified 66 dense cores in the Ophiuchus molecular cloud.
We apply a simple definition that protostellar dense cores are separated
by no more than 15\arcsec\ (i.e., roughly one SCUBA 850~\mum\ beam) from one (or more)
{\it Spitzer}-identified YSOs.  Using this requirement, we 
conservatively classify 43 cores as starless 
and 23 cores as protostellar.
We use the updated {\it Spitzer} YSO catalogue of \citet{Dunham15} for this determination,
and further note that of the 23 protostellar cores, 14 are associated with
a Class 0+I YSO, 4 are associated with a Class `Flat' YSO, and 5 are associated with
a Class II YSO.  These classifications are based on the extinction-corrected slope
of the spectral energy distribution measured across the {\it Spitzer} bands
\citep[$\alpha_0$ from][]{Dunham15}.

None of the 43 starless cores in our list contain candidate embedded YSOs in
\citet{Jorgensen08} or are included in the candidate faint embedded YSO list of 
\citet{Dunham08}.  
Two of the starless cores in our list (162628-24235 and 162628-24225), however, 
match cores that \citet{Sadavoy10a} identify as having evidence of a protostellar nature.

\begin{figure}[htb!]
\includegraphics[width=6in]{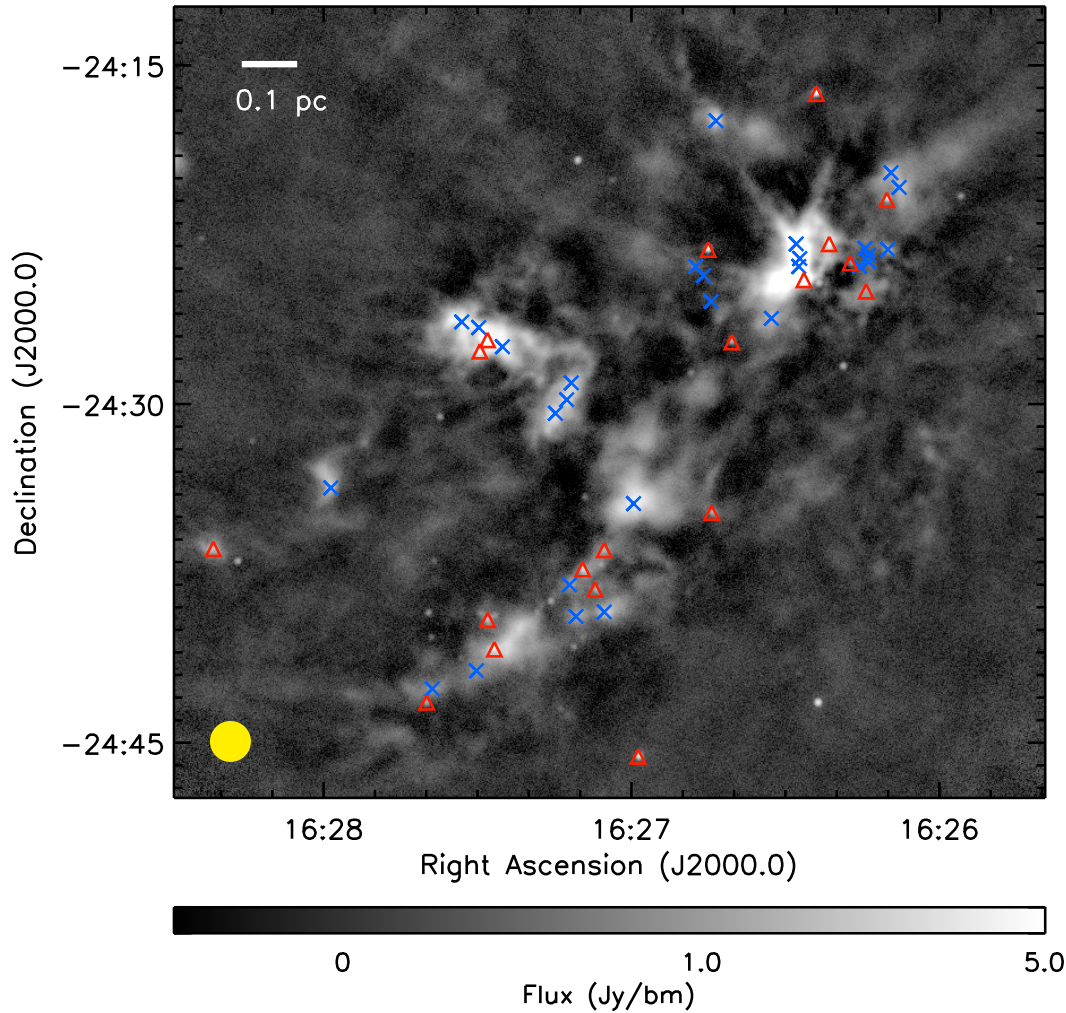} 
\caption{The location of the dense cores targeted in our ALMA observations in
	the western portion of Ophiuchus.
	The background greyscale image shows the 
	SCUBA-2 850~\mum\ map created using the JCMT GBS Data Release~2 
	reduction method (Kirk et.~al.,~in prep).  
	The blue crosses show starless dense cores, while
	the red triangles show protostellar dense cores.
	The filled yellow circle at the bottom left corner indicates the 
	size of
	the ALMA primary beam.
	The horizontal line in the upper left corner indicates a length of
	0.1~pc at the assumed distance to Ophiuchus of 140~pc. 
	}
\label{fig_scuba2_posns}
\end{figure}

\begin{figure}[htbp!]
\includegraphics[width=5.5in]{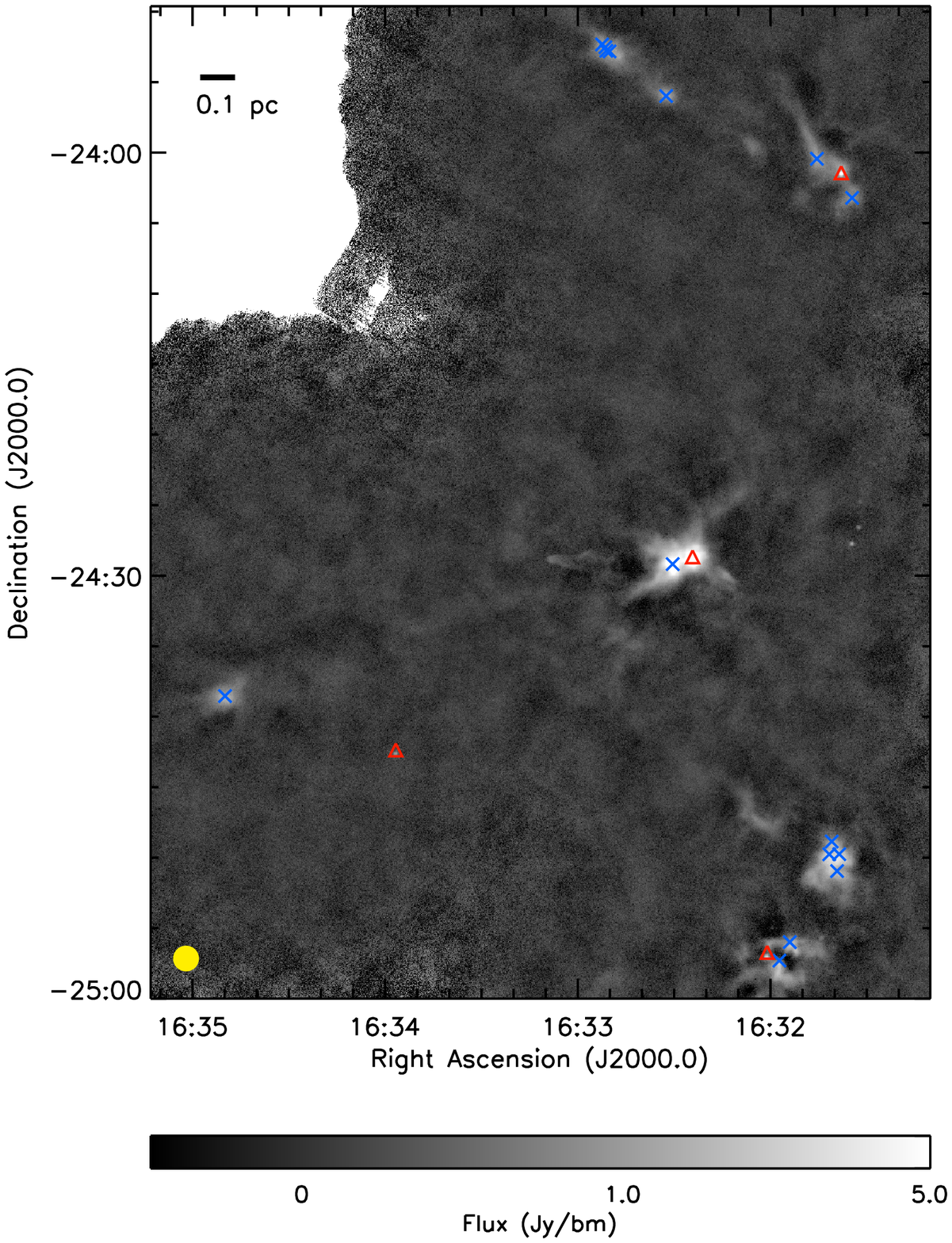}
\caption{The location of the dense cores targeted in our ALMA observations in
	the eastern portion of Ophiuchus.
	The background greyscale image shows the 
	SCUBA-2 850~\mum\ map created using the JCMT GBS Data Release~2 
	reduction method (Kirk et al., in prep).  
	The blue crosses show starless dense cores, while
	the red triangles show protostellar dense cores.
	The filled yellow circle at the bottom left corner indicates the 
	size of
	the ALMA primary beam.
	The horizontal line in the upper left corner indicates a length of
	0.1~pc at the assumed distance to Ophiuchus of 140~pc. 
	}
\label{fig_scuba2_posns2}
\end{figure}

We obtained ALMA observations for all 66 of the dense cores identified in \citet{Jorgensen08}.
Figures~\ref{fig_scuba2_posns} and \ref{fig_scuba2_posns2} show the positions of the 66 
pointing centres overlaid on the recent SCUBA-2 850~\mum\ map of Ophiuchus from \citet{Pattle15}.


\subsubsection{Comparison of SCUBA and SCUBA-2 data}
Although not available when the ALMA observations were proposed, the Ophiuchus cloud was recently 
observed and analyzed by the JCMT Gould Belt Survey by
\citet{Pattle15}, which has larger spatial coverage and better flux sensitivity than the
original SCUBA observations used to create the ALMA target list.  
We examined our adopted dense core catalogue using this
improved SCUBA-2 dataset.  Unfortunately, six of the cores 
identified by \citeauthor{Jorgensen08} appear to be spurious
objects: 162614-24232, 162616-24235, 162644-24253, 162646-24242, 163248-23523, and 163248-23524.
All six appear to be part of larger structures which were identified as other 
sources in the 
\citet{Jorgensen08} catalogue.  Examining the original SCUBA maps, it appears that
a combination of poor larger-scale flux recovery and a higher local pixel-to-pixel noise
caused the core identifying algorithm they used in their analysis 
\citep[{\it clumpfind};][]{Williams94} to
subdivide the emission artificially into multiple structures.  
\citep[Challenges with using
{\it clumpfind} are discussed in detail in][]{Pineda09}.  Since all six sources 
are the result of artificial subdivision of larger emission structures which were
also included in our catalogue and observed, the ALMA observations serve to provide
deeper integrations on those adjacent real dense cores.
All six spurious sources fall into our starless core category, giving us a total
of 37 starless cores and 23 protostellar cores.
For the remainder of our analysis, we therefore consider our observational sample
to include 60 cores, and use the data from the six spurious cores only to improve
the sensitivity toward their neighbouring real cores.

We furthermore compared the SCUBA-based core catalogue with the core catalogue presented
in \citet{Pattle15} using SCUBA-2 observations.  \citet{Pattle15} used the core-finding
algorithm CuTEx \citep{Molinari11}, which uses the second derivative of the emission map to 
identify objects.  In practice, this approach 
renders CuTEx insensitive to larger, fluffier sources, 
while giving it an advantage in breaking up complex clustered emission into multiple sources.
As such, we find that some of the fainter SCUBA cores 
are not listed in
the \citet{Pattle15}
catalogue, despite there being visually obvious emission at their 
locations.  Moreover, other SCUBA
cores are associated with multiple listings 
in the \citet{Pattle15} catalogue, as CuTEx subdivided
the respective emission to a much greater extent than {\it clumpfind}
did.  We emphasize that in general, there is good agreement
between the two catalogues for the brightest cores, while fainter cores have more likelihood
to differ for the reasons outlined above.  
Since there is not a 
one-to-one correspondence between the two catalogues, for the remainder of our
analysis we present the properties derived for the cores from the SCUBA data in
\citet{Jorgensen08}.

\subsection{ALMA Data}

The ALMA Band 3 data were observed during Cycle 2 on 31 January, 2015 with 37 antennas.  
The array was
in a relatively compact configuration, with a median synthesized beam of 
$3.5 \arcsec \times 1.8\arcsec $ at a position angle of 71$^o$, and a 
maximum recoverable scale of $\sim$23\arcsec.  The former corresponds to a physical
size of $\sim$0.002~pc or 350~AU at a distance of 140~pc, while the latter corresponds
to $\sim$0.016~pc.
Our observational setup mimics that described in
\citetalias{Dunham16} for their earlier observations in Chamaeleon~I.  
Only the 12~m array was used, i.e., no compact array or total power 
observations were included.
Antenna separations ranged from $\sim$15~m to 350~m.
Each core was observed twice with
a single pointing using roughly 1 minute of on-source integration time.  Three
of the four spectral windows available were configured for continuum measurements
at 101~GHz, 103~GHz, and 113~GHz, each covering a bandwidth of 2~GHz, and giving an effective mean
frequency of 107~GHz.  The noise
level of the continuum observations about the phase centres is roughly 0.16~mJy~beam$^{-1}$.
The remaining spectral window was configured to observe $^{12}$CO~(1-0) emission at 115~GHz;
the line data are not presented here.  

We used the pipeline calibrated data available for Cycle 2 data
\citep{Shinnaga15}. 
For fields where a bright enough source was detected, we manually self-calibrated the
data in  both phase and amplitude
using the Common Astronomy Software Applications (CASA) package, and
we then imaged the data using manually defined clean boxes in an interactive
clean session.  In nearly all cases, self-calibration gave only a 
modest (ten percent or less) improvement on the rms noise level in the final image.
Some of the observed cores lie sufficiently close together for
their fields of view to overlap.  In these cases, we mosaicked the fields together to
improve the final sensitivity.
Our observation of core 162628-24235 was omitted from our analysis, as it
suffered from a known bug at ALMA, and gave inconsistent results.
The area of core 162628-24235 was covered by
our ALMA observations of neighbouring cores in the mosaic 162628-24225, so the exclusion
of this one observation does not impact our science goals.
Table~\ref{tab_rms} gives the final noise level for each core observed, along with our
classification as starless or protostellar.  Those cores 
lying within a mosaicked area are noted.

\startlongtable
\begin{deluxetable}{ccccc}
\tablecolumns{5}
\tablewidth{0pt}
\tabletypesize{\scriptsize}
\tablecaption{Noise levels of ALMA observations\label{tab_rms}}
\tablehead{
\colhead{Field\tablenotemark{a}} &
\colhead{rms (mJy~beam$^{-1}$)\tablenotemark{b}} &
\colhead{Mosaic Field?\tablenotemark{c}} &
\colhead{Protostellar?\tablenotemark{d}} &
\colhead{Concentration\tablenotemark{e}} 
}
\startdata
 162608-24202 & 0.155 & 162610-24206 &     S &  0.34 \\
 162610-24195 & 0.155 & 162610-24206 &     S &  0.34 \\
 162610-24231 & 0.164 &           -- &     S &  0.40 \\
 162610-24206 & 0.155 & 162610-24206 &    II &  0.65 \\
 162614-24232* & 0.129 & 162617-24235 &     S &  0.15 \\
 162614-24234 & 0.129 & 162617-24235 &     S &  0.22 \\
 162614-24250 & 0.165 &           -- &   0+I &  0.34 \\
 162615-24231 & 0.129 & 162617-24235 &     S &  0.17 \\
 162616-24235* & 0.129 & 162617-24235 &     S &  0.20 \\
 162617-24235 & 0.129 & 162617-24235 &   0+I &  0.32 \\
 162622-24225 & 0.159 &           -- &   0+I &  0.58 \\
 162624-24162 & 0.172 &           -- &    II &  0.41 \\
 162626-24243 & 0.167 & 162628-24225 &   0+I &  0.80 \\
 162627-24233 & 0.167 & 162628-24225 &     S &  0.66 \\
 162628-24235 & 0.167 & 162628-24225 &     S &  0.80 \\
 162628-24225 & 0.167 & 162628-24225 &     S &  0.66 \\
 162633-24261 & 0.162 &           -- &     S &  0.41 \\
 162641-24272 & 0.156 &           -- &   0+I &  0.53 \\
 162644-24173 & 0.166 &           -- &     S &  0.40 \\
 162644-24345 & 0.170 &           -- &   0+I &  0.29 \\
 162644-24253* & 0.168 &           -- &     S &  0.23 \\
 162645-24231 & 0.159 &           -- &    II &  0.74 \\
 162646-24242* & 0.114 & 162646-24242 &     S &  0.16 \\
 162648-24236 & 0.114 & 162646-24242 &     S &  0.33 \\
 162659-24454 & 0.163 &           -- &    II &  0.42 \\
 162660-24343 & 0.164 &           -- &     S &  0.55 \\
 162705-24363 & 0.155 &           -- &   0+I &  0.34 \\
 162705-24391 & 0.168 &           -- &     S &  0.44 \\
 162707-24381 & 0.171 & 162709-24372 &   0+I &  0.22 \\
 162709-24372 & 0.171 & 162709-24372 &   0+I &  0.48 \\
 162711-24393 & 0.160 &           -- &     S &  0.29 \\
 162712-24290 & 0.164 &           -- &     S &  0.21 \\
 162712-24380 & 0.145 &           -- &     S &  0.31 \\
 162713-24295 & 0.158 &           -- &     S &  0.38 \\
 162715-24303 & 0.157 &           -- &     S &  0.34 \\
 162725-24273 & 0.146 & 162730-24264 &     S &  0.37 \\
 162727-24405 & 0.158 &           -- &   0+I &  0.57 \\
 162728-24271 & 0.146 & 162730-24264 &  Flat &  0.34 \\
 162728-24393 & 0.161 &           -- &   0+I &  0.37 \\
 162729-24274 & 0.146 & 162730-24264 &  Flat &  0.50 \\
 162730-24264 & 0.146 & 162730-24264 &     S &  0.35 \\
 162730-24415 & 0.158 &           -- &     S &  0.32 \\
 162733-24262 & 0.164 &           -- &     S &  0.45 \\
 162739-24424 & 0.150 & 162740-24431 &     S &  0.31 \\
 162740-24431 & 0.150 & 162740-24431 &  Flat &  0.27 \\
 162759-24334 & 0.141 &           -- &     S &  0.46 \\
 162821-24362 & 0.155 &           -- &   0+I &  0.39 \\
 163133-24032 & 0.165 &           -- &     S &  0.35 \\
 163136-24013 & 0.149 &           -- &  Flat &  0.72 \\
 163138-24495 & 0.152 &           -- &     S &  0.38 \\
 163139-24506 & 0.157 &           -- &     S &  0.35 \\
 163140-24485 & 0.163 &           -- &     S &  0.29 \\
 163141-24495 & 0.163 &           -- &     S &  0.36 \\
 163143-24003 & 0.164 &           -- &     S &  0.33 \\
 163154-24560 & 0.172 &           -- &     S &  0.40 \\
 163157-24572 & 0.159 &           -- &     S &  0.46 \\
 163201-24564 & 0.155 &           -- &   0+I &  0.45 \\
 163223-24284 & 0.827 &           -- &   0+I &  0.94 \\
 163229-24291 & 0.167 &           -- &     S &  0.74 \\
 163230-23556 & 0.166 &           -- &     S &  0.33 \\
 163247-23524 & 0.097 & 163249-23521 &     S &  0.26 \\
 163248-23524* & 0.097 & 163249-23521 &     S &  0.30 \\
 163248-23523* & 0.097 & 163249-23521 &     S &  0.19 \\
 163249-23521 & 0.097 & 163249-23521 &     S &  0.28 \\
 163356-24420 & 0.177 &           -- &    II &  0.33 \\
 163448-24381 & 0.147 &           -- &     S &  0.38 \\
\enddata
\tablenotetext{a}{Observed SCUBA core name from \citet{Jorgensen08}.  Asterisks denote the
	six starless cores from that sample which more sensitive SCUBA-2 observations show
	were likely false detections.  See text for more details.}
\tablenotetext{b}{Typical map / mosaic rms value for areas with no detections calculated
	using CASA's imstat command.  The rms was calculated on the non-primary-beam
	corrected image, ensuring that representative noise values could be accurately
	measured even for images where there was a bright central point source.}
\tablenotetext{c}{For areas with multiple SCUBA cores having contiguous ALMA coverage, the
	eastern most core in the covered zone is listed.  Dashes denote isolated fields.}
\tablenotetext{d}{Protostellar class based on the {\it Spitzer} catalogue of \citet{Dunham15}.
	`S' denotes starless cores.}
\tablenotetext{e}{Central concentration of each core as reported in \citet{Jorgensen08}.}
\end{deluxetable}

\section{Detections}

After correcting for the primary beam, we use CASA's {\it imfit} package to fit 
two-dimensional Gaussians to all compact sources of emission within the single maps and 
mosaicked areas.  
Table~\ref{tab_detects_pt1} lists the the 38 sources detected
with ALMA and the association of each with
the nearest SCUBA core (and separation), and nearest {\it Spitzer} YSO (and separation).
Sources which do not have a close positional association with a {\it Spitzer} YSO, but
have other information suggesting a protostellar nature (see Section~3.3) are 
also noted.

The results of the Gaussian fits to each ALMA detection are given in
Table~\ref{tab_detects_pt2}.  
Where possible, we calculate the beam-deconvolved size of the source.  Marginally
resolved sources are noted in Table~\ref{tab_detects_pt2} with upper limit signs
for the
deconvolved sizes and no position angles.  
Unresolved sources have no value listed
for either a deconvolved size or position angle.

\begin{deluxetable}{cccccccc}
\tablecolumns{8}
\tablewidth{0pt}
\tabletypesize{\scriptsize}
\tablecaption{ALMA Detections and Associations\label{tab_detects_pt1}}
\tablehead{
\colhead{Src\tablenotemark{a}} &
\colhead{R.A.\tablenotemark{a}} &
\colhead{Decl.\tablenotemark{a}} &
\colhead{SCUBA\tablenotemark{b}} &
\colhead{Sep.\tablenotemark{b}} &
\colhead{{\it Spitzer}\tablenotemark{c}}&
\colhead{Sep.\tablenotemark{c}} &
\colhead{Common\tablenotemark{d}}
\\
\colhead{\#} &
\colhead{(J2000)} &
\colhead{(J2000)} &
\colhead{Core} &
\colhead{(\arcsec)} &
\colhead{Match?} &
\colhead{(\arcsec)} &
\colhead{Name} 
}
\startdata
   1 & 16:26:10.33 & -24:20:55.17 &162610-24206 &   2.6 &    J162610.3-242054 &   1.2 &              ... \\
   2 & 16:26:17.24 & -24:23:45.72 &162617-24235 &   3.7 &    J162617.2-242345 &   0.9 &              ... \\
   3 & 16:26:18.98 & -24:24:14.69 &162617-24235 &  35.3 &    J162618.9-242414 &   1.2 &              ... \\
   4 & 16:26:21.39 & -24:23:04.78 &162622-24225 &  10.4 &    J162621.3-242304 &   1.4 &              ... \\
   5 & 16:26:21.73 & -24:22:50.93 &162622-24225 &   4.3 &             Hersch. &  14.3 &              ... \\
   6 & 16:26:24.09 & -24:16:13.81 &162624-24162 &   2.1 &    J162624.0-241613 &   1.5 &              ... \\
   7 & 16:26:25.49 & -24:23:01.75 &162628-24225 &  34.9 &    J162625.4-242301 &   1.5 &              ... \\
   8 & 16:26:25.63 & -24:24:29.54 &162626-24243 &  11.9 &    J162625.6-242428 &   1.6 &        VLA~1623W \\
   9 & 16:26:26.36 & -24:24:30.72 &162626-24243 &   1.9 &    J162626.4-242430 &   0.9 &         VLA~1623 \\
  10 & 16:26:27.43 & -24:24:18.26 &162626-24243 &  17.9 &    X-ray + Hersch.? &  18.3 &              ... \\
  11 & 16:26:27.86 & -24:23:59.56 &162628-24235 &   6.7 &    X-ray + Hersch.? &  33.1 &         OphA-SM1 \\
  12 & 16:26:30.48 & -24:22:57.55 &162628-24225 &  34.0 &c2d-uncut + Hersch.? &  69.5 &              ... \\
  13 & 16:26:40.47 & -24:27:14.90 &162641-24272 &   1.3 &    J162640.4-242714 &   1.3 &              ... \\
  14 & 16:26:44.21 & -24:34:48.81 &162644-24345 &   2.7 &    J162644.1-243448 &   1.6 &              ... \\
  15 & 16:26:45.02 & -24:23:08.08 &162645-24231 &   2.4 &    J162645.0-242307 &   1.1 &              ... \\
  16 & 16:26:58.50 & -24:45:36.96 &162659-24454 &   2.7 &             Hersch. &   6.1 &              ... \\
  17 & 16:26:59.15 & -24:34:58.99 &162660-24343 &  34.2 &             Hersch. &   4.1 &              ... \\
  18 & 16:27:05.25 & -24:36:29.90 &162705-24363 &   1.7 &    J162705.2-243629 &   1.1 &              ... \\
  19 & 16:27:05.86 & -24:37:08.36 &162705-24363 &  40.7 &             Hersch. &  40.4 &              ... \\
  20 & 16:27:06.76 & -24:38:15.21 &162707-24381 &   4.9 &    J162706.7-243814 &   1.5 &              ... \\
  21 & 16:27:09.41 & -24:37:18.96 &162709-24372 &   1.3 &    J162709.4-243718 &   1.0 &              ... \\
  22 & 16:27:09.49 & -24:37:22.89 &162709-24372 &   3.5 &             Hersch. &   5.0 &              ... \\
  23 & 16:27:26.47 & -24:39:23.42 &162728-24393 &  23.5 &     X-ray + Hersch. &  21.6 &              ... \\
  24 & 16:27:26.60 & -24:40:45.51 &162727-24405 &   6.9 &             Hersch. &   6.2 &              ... \\
  25 & 16:27:26.91 & -24:40:50.71 &162727-24405 &   3.4 &    J162726.9-244050 &   0.7 &              ... \\
  26 & 16:27:28.00 & -24:39:33.76 &162728-24393 &   0.5 &    J162727.9-243933 &   1.6 &              ... \\
  27 & 16:27:29.44 & -24:39:16.12 &162728-24393 &  26.8 &             Hersch. &  27.0 &              ... \\
  28 & 16:27:30.18 & -24:27:43.74 &162729-24274 &   9.9 &    J162730.1-242743 &   1.3 &              ... \\
  29 & 16:27:37.22 & -24:42:38.51 &162739-24424 &  21.6 &    J162737.2-244237 &   1.5 &              ... \\
  30 & 16:27:39.82 & -24:43:15.22 &162740-24431 &   2.3 &    J162739.8-244315 &   0.4 &              ... \\
  31 & 16:28:21.63 & -24:36:23.78 &162821-24362 &   3.2 &    J162821.6-243623 &   0.9 &  IRAS~16253-2429 \\
  32 & 16:31:35.66 & -24:01:29.78 &163136-24013 &   2.6 &    J163135.6-240129 &   1.2 &              ... \\
  33 & 16:31:52.11 & -24:56:15.99 &163154-24560 &  27.9 &    J163152.1-245615 &   1.0 &              ... \\
  34 & 16:32:00.98 & -24:56:43.14 &163201-24564 &   1.3 &    J163200.9-245642 &   1.6 &              ... \\
  35 & 16:32:22.63 & -24:28:32.26 &163223-24284 &   6.1 &    J163222.6-242831 &   1.3 &  IRAS~16293-2422 \\
  36 & 16:32:22.89 & -24:28:36.26 &163223-24284 &   0.9 &             Hersch. &   6.6 &  IRAS~16293-2422 \\
  37 & 16:32:29.06 & -24:29:10.27 &163229-24291 &   3.1 &                 ... &  96.6 &              ... \\
  38 & 16:33:55.63 & -24:42:05.50 &163356-24420 &   1.6 &    J163355.6-244205 &   0.6 &              ... \\
\enddata
\tablenotemark{a}{Running index number and peak position of the ALMA detection.}
\tablenotetext{b}{Closest SCUBA core from \citet{Jorgensen08}, and the angular separation between
	the SCUBA core centre and the ALMA detection's peak position.}
\tablenotetext{c}{Protostellar associations.  Angular separations to the nearest 
	{\it Spitzer} YSO in \citet{Dunham15} are given.  
	The \citet{Dunham15} source names are
	listed for sources with separations of less than 3\arcsec.
	Associations with {\it Herschel} emission suggesting the presence of a protostar,
	or an X-ray candidate YSO, or a YSO candidate in the original {\it Spitzer} c2d catalogue 
	are listed as appropriate for sources without a close
	\citet{Dunham15} {\it Spitzer} match.}
\tablenotetext{d}{Commonly known protostellar associations, where applicable.}
\end{deluxetable}

\begin{deluxetable}{ccccccccccccc}
\tablecolumns{13}
\tablewidth{0pt}
\tabletypesize{\scriptsize}
\tablecaption{Observed Properties of ALMA Detections\label{tab_detects_pt2}}
\tablehead{
\colhead{Src} &
\colhead{Peak\tablenotemark{a}}&
\colhead{P$_{err}$\tablenotemark{a}}&
\colhead{Total\tablenotemark{a}} &
\colhead{T$_{err}$\tablenotemark{a}} &
\colhead{FWHM$_a$\tablenotemark{a}} &
\colhead{FWHM$_b$\tablenotemark{a}} &
\colhead{P.A.\tablenotemark{a}} &
\colhead{lim.\tablenotemark{b}} &
\colhead{FWHM$_{a,d}$\tablenotemark{b}} &
\colhead{FWHM$_{b,d}$\tablenotemark{b}} &
\colhead{P.A.$_{d}$\tablenotemark{b}} & 
\colhead{P.A.$_{d,e}$\tablenotemark{b}} 
\\
\colhead{\#} &
\multicolumn{2}{c}{(mJy~beam$^{-1}$)} &
\multicolumn{2}{c}{(mJy)}&
\colhead{(arcsec)} &
\colhead{(arcsec)} &
\colhead{(deg)}&
\colhead{} &
\colhead{(arcsec)} &
\colhead{(arcsec)} &
\colhead{(deg)} &
\colhead{(deg)}
}
\startdata
  1 &  27.25 &   0.35 &  27.69 &   0.65 &  3.613 &  1.855 &  71.2 & $<$ &  0.70 &  0.40 &  -1 &  -1 \\
  2 &  14.46 &   0.29 &  15.19 &   0.57 &  3.712 &  1.887 &  71.1 & ... &  0.74 &  0.42 &  53 &  21 \\
  3 &   5.96 &   0.28 &   5.44 &   0.50 &  3.634 &  1.673 &  75.8 & $<$ & -1.00 & -1.00 &  -1 &  -1 \\
  4 &   3.48 &   0.67 &   4.30 &   1.40 &  4.240 &  2.400 &  77.0 & ... &  1.80 &  0.80 & 107 &  44 \\
  5 &  30.71 &   0.63 &  31.00 &   1.20 &  3.928 &  2.104 &  71.3 & $<$ &  1.00 &  0.40 &  -1 &  -1 \\
  6 &  49.25 &   0.73 &  52.00 &   1.40 &  3.990 &  2.131 &  71.6 & ... &  0.66 &  0.51 &  14 &  65 \\
  7 &   9.17 &   0.44 &   8.95 &   0.80 &  3.452 &  1.855 &  66.1 & $<$ & -1.00 & -1.00 &  -1 &  -1 \\
  8 &  11.24 &   0.44 &  11.24 &   0.82 &  3.547 &  1.850 &  69.2 & $<$ & -1.00 & -1.00 &  -1 &  -1 \\
  9 &  59.82 &   0.47 &  78.70 &   1.00 &  3.948 &  2.186 &  73.1 & ... &  1.68 &  1.15 &  87 &   6 \\
 10 &   7.10 &   0.44 &   7.56 &   0.86 &  3.744 &  1.866 &  71.3 & ... &  1.08 &  0.36 &  76 &  12 \\
 11 &  23.13 &   0.46 &  28.75 &   0.98 &  3.913 &  2.084 &  73.5 & ... &  1.61 &  0.93 &  89 &  17 \\
 12 &  11.22 &   0.44 &  11.88 &   0.86 &  3.736 &  1.860 &  69.6 & ... &  1.07 &  0.25 &  58 &  14 \\
 13 &  10.95 &   0.15 &  12.04 &   0.30 &  3.939 &  2.182 &  73.2 & ... &  0.92 &  0.47 & 137 &  45 \\
 14 &  16.83 &   0.66 &  16.90 &   1.20 &  3.428 &  1.749 &  70.6 & $<$ &  1.10 &  0.40 &  -1 &  -1 \\
 15 &  36.27 &   0.67 &  43.40 &   1.40 &  4.001 &  2.379 &  76.4 & ... &  1.42 &  0.61 & 129 &  19 \\
 16 &  29.64 &   0.67 &  31.20 &   1.30 &  3.830 &  2.138 &  70.4 & ... &  0.69 &  0.34 &   9 &  66 \\
 17 &   7.22 &   0.39 &   7.97 &   0.75 &  3.492 &  1.907 &  62.3 & $<$ & -1.00 & -1.00 &  -1 &  -1 \\
 18 &   9.54 &   0.64 &   9.80 &   1.20 &  3.740 &  2.110 &  72.4 & $<$ &  1.60 &  0.80 &  -1 &  -1 \\
 19 &   8.20 &   0.63 &   9.20 &   1.30 &  4.230 &  2.040 &  76.6 & $<$ &  2.90 &  0.70 &  -1 &  -1 \\
 20 &   9.87 &   0.13 &   9.92 &   0.25 &  3.612 &  1.804 &  72.8 & $<$ &  0.90 &  0.20 &  -1 &  -1 \\
 21 &   4.47 &   0.14 &   5.00 &   0.28 &  3.649 &  1.986 &  70.7 & ... &  0.94 &  0.52 &  19 &  51 \\
 22 &   0.61 &   0.12 &   5.10 &   1.10 & 11.050 &  4.900 &  63.4 & ... & 10.50 &  4.50 &  62 &  11 \\
 23 &   8.38 &   0.65 &   8.50 &   1.20 &  3.427 &  1.749 &  71.6 & ... &  0.56 &  0.17 &  81 &  21 \\
 24 &   3.07 &   0.57 &   2.64 &   0.99 &  3.280 &  1.550 &  73.0 & $<$ & -1.00 & -1.00 &  -1 &  -1 \\
 25 &   5.56 &   0.64 &   6.80 &   1.30 &  3.640 &  1.980 &  76.7 & ... &  1.52 &  0.62 & 105 &  33 \\
 26 &   4.34 &   0.68 &   4.90 &   1.30 &  3.480 &  1.900 &  72.0 & ... &  0.86 &  0.73 & 108 &  55 \\
 27 &   2.31 &   0.65 &   2.70 &   1.40 &  3.940 &  1.770 &  76.6 & $<$ &  4.50 &  1.20 &  -1 &  -1 \\
 28 &   2.49 &   0.13 &   2.67 &   0.25 &  3.602 &  1.908 &  68.1 & $<$ &  1.70 &  0.80 &  -1 &  -1 \\
 29 &   1.39 &   0.36 &   1.43 &   0.66 &  3.290 &  1.910 &  33.0 & $<$ & -1.00 & -1.00 &  -1 &  -1 \\
 30 &  15.88 &   0.36 &  17.06 &   0.68 &  3.444 &  1.903 &  70.6 & ... &  0.74 &  0.35 &  15 &  53 \\
 31 &   2.35 &   0.15 &   3.05 &   0.32 &  3.835 &  1.955 &  72.3 & ... &  1.89 &  0.89 &  75 &  46 \\
 32 &  50.77 &   0.61 &  51.80 &   1.10 &  3.770 &  2.108 &  72.8 & $<$ &  0.70 &  0.50 &  -1 &  -1 \\
 33 &   9.22 &   0.69 &  10.60 &   1.40 &  3.842 &  1.752 &  67.8 & $<$ &  2.70 &  0.60 &  -1 &  -1 \\
 34 &  12.23 &   0.20 &  14.24 &   0.40 &  4.041 &  2.294 &  67.6 & ... &  1.70 &  0.60 &  50 &   9 \\
 35 & 294.60 &   3.70 & 328.10 &   7.20 &  4.065 &  2.205 &  73.1 & ... &  1.33 &  0.67 &  81 &  14 \\
 36 & 129.60 &   4.10 & 246.00 &  11.00 &  4.506 &  3.386 &  58.1 & ... &  2.96 &  1.97 &  11 &  15 \\
 37 &   0.70 &   0.14 &   7.30 &   1.60 & 10.300 &  6.000 &   4.8 & ... & 10.10 &  5.10 &   2 &  14 \\
 38 &  11.13 &   0.14 &  10.90 &   0.26 &  3.836 &  2.089 &  72.1 & $<$ &  0.30 &  0.10 &  -1 &  -1 \\
\enddata
\tablenotetext{a}{Properties of the best-fit Gaussian to the ALMA emission: 
	peak flux, total flux, FWHM of the major and minor axes, and position angle.}
\tablenotetext{b}{Deconvolved size estimates of the best-fit Gaussian: whether the estimate is
	a lower limit, FWHM of the major and minor axes, and position angle and associated error.  
	Unresolved sources have FWHM and position angle values of -1.}
\end{deluxetable}

\subsection{Mass Estimates}
We can estimate the mass of each detected ALMA source using the standard equation
\begin{equation}
M = 100 \frac{d^2S_{\nu}}{B_{\nu}(T_D)\kappa{_\nu}}
\label{eqn_mass}
\end{equation}
where $d$ is the distance, $S_{\nu}$ is the flux at frequency $\nu$, $B_{\nu}$ is the
Planck function at a dust temperature of $T_D$, $\kappa_{\nu}$ is the dust opacity,
and the factor of 100 represents the assumed gas-to-dust ratio.
Following \citetalias{Dunham16}, we assume $\kappa_{\nu} = 0.23$~cm~$^2$g$^{-1}$ at an
effective frequency of 107~GHz, which corresponds to the OH5 model of \citet{Ossenkopf94}.  
We assume a slightly warmer dust
temperature for Ophiuchus than \citetalias{Dunham16} assumed for Chamaeleon~I
(15~K versus 10~K), based on the
NH$_3$ results of \citet{Friesen09},
and a distance of 140~pc.  
Note that on the small scales of the 
densest gas traced by ALMA, the temperatures of starless cores (i.e., source 37
and the extended emission discussed in Sections~3.4 and 3.5) are likely closer to a temperature
of 10~K, as discussed in Section~5.  A temperature of 10~K would imply masses that are a 
factor of two higher than what we report.  Conversely, the protostellar cores may be warmer
on small size
scales.  A temperature of 25~K would imply true masses that are roughly a factor of two lower
than what we report.  The temperature and dust opacity contribute a factor of several to
the uncertainty in the masses we estimate, while uncertainty in the distance has a smaller
contribution.

Table~\ref{tab_mass} summarizes the estimated mass of each source detected with ALMA,
in addition to their effective radius and mean density.  We calculate the effective radius
as the geometric mean of the deconvolved semi-major and semi-minor 
axes (i.e., half of the reported FWHM
values reported in Table~\ref{tab_detects_pt2}).  In the case of unresolved sources, we use
the (clean) beam major and minor axes as an upper limit.  Following \citetalias{Dunham16},
we calculate the mean density of each source as
\begin{equation}
n = \frac{3}{4\pi \mu m_H } \frac{M}{R_{eff}^3}
\label{eqn_dens}
\end{equation}

where $\mu = 2.37$ is the mean molecular weight per particle and $m_H$ is the mass of 
a hydrogen atom.

\begin{deluxetable}{cccccc}
\tablecolumns{6}
\tablewidth{0pt}
\tabletypesize{\scriptsize}
\tablecaption{Physical Properties of Detections\label{tab_mass}}
\tablehead{
\colhead{Src\tablenotemark{a}} &
\colhead{Mass\tablenotemark{b}} &
\colhead{} &
\colhead{R$_{eff}$\tablenotemark{b}} &
\colhead{}&
\colhead{n\tablenotemark{b}}
\\
\colhead{Num} &
\colhead{(M$_{\odot}$)} &
\colhead{} &
\colhead{(au)} &
\colhead{} &
\colhead{(cm$^{-3}$)}
}
\startdata
   1 &  0.2602 & $<$ &  37 & $>$ &  1.83E+11 \\
   2 &  0.1427 & ... &  39 & ... &  8.61E+10 \\
   3 &  0.0511 & $<$ & 788 & $>$ &  3.74E+06 \\
   4 &  0.0404 & ... &  83 & ... &  2.44E+09 \\
   5 &  0.2913 & $<$ &  44 & $>$ &  1.20E+11 \\
   6 &  0.4885 & ... &  40 & ... &  2.61E+11 \\
   7 &  0.0841 & $<$ & 775 & $>$ &  6.47E+06 \\
   8 &  0.1056 & $<$ & 792 & $>$ &  7.62E+06 \\
   9 &  0.7394 & ... &  97 & ... &  2.88E+10 \\
  10 &  0.0710 & ... &  43 & ... &  3.06E+10 \\
  11 &  0.2701 & ... &  85 & ... &  1.54E+10 \\
  12 &  0.1116 & ... &  36 & ... &  8.43E+10 \\
  13 &  0.1131 & ... &  46 & ... &  4.16E+10 \\
  14 &  0.1588 & $<$ &  46 & $>$ &  5.68E+10 \\
  15 &  0.4078 & ... &  65 & ... &  5.29E+10 \\
  16 &  0.2931 & ... &  33 & ... &  2.70E+11 \\
  17 &  0.0749 & $<$ & 762 & $>$ &  6.04E+06 \\
  18 &  0.0921 & $<$ &  79 & $>$ &  6.64E+09 \\
  19 &  0.0864 & $<$ &  99 & $>$ &  3.12E+09 \\
  20 &  0.0932 & $<$ &  29 & $>$ &  1.28E+11 \\
  21 &  0.0470 & ... &  48 & ... &  1.44E+10 \\
  22 &  0.0479 & ... & 481 & ... &  1.54E+07 \\
  23 &  0.0799 & ... &  21 & ... &  2.84E+11 \\
  24 &  0.0248 & $<$ & 744 & $>$ &  2.15E+06 \\
  25 &  0.0639 & ... &  67 & ... &  7.30E+09 \\
  26 &  0.0460 & ... &  55 & ... &  9.67E+09 \\
  27 &  0.0254 & $<$ & 162 & $>$ &  2.11E+08 \\
  28 &  0.0251 & $<$ &  81 & $>$ &  1.65E+09 \\
  29 &  0.0134 & $<$ & 555 & $>$ &  2.81E+06 \\
  30 &  0.1603 & ... &  35 & ... &  1.27E+11 \\
  31 &  0.0287 & ... &  90 & ... &  1.37E+09 \\
  32 &  0.4867 & $<$ &  41 & $>$ &  2.46E+11 \\
  33 &  0.0996 & $<$ &  89 & $>$ &  5.05E+09 \\
  34 &  0.1338 & ... &  70 & ... &  1.36E+10 \\
  35 &  3.0826 & ... &  66 & ... &  3.83E+11 \\
  36 &  2.3112 & ... & 169 & ... &  1.72E+10 \\
  37 &  0.0686 & ... & 502 & ... &  1.94E+07 \\
  38 &  0.1024 & $<$ &  12 & $>$ &  2.06E+12 \\
\enddata
\tablenotetext{a}{Source number, from Table~\ref{tab_detects_pt1}.}
\tablenotetext{b}{Estimated mass, effective radius (or upper limit), and 
	density (or lower limit) for all ALMA detections.  See text for a 
	discussion on the sources of uncertainty in the mass estimates.}
\end{deluxetable}

For all quantities reported in Table~\ref{tab_mass}, the uncertainty is dominated by
systematic effects (distance, temperature, dust opacity, and source geometry), 
rather than the statistical uncertainties associated with each fit.

\subsection{Detections and Protostellar Associations}
\begin{figure}[htb]
\includegraphics[width=4in]{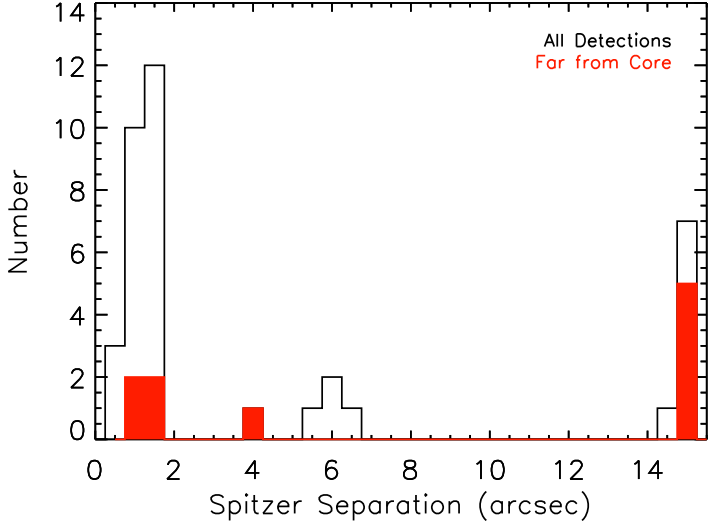}
\caption{
	Distribution of angular separations between ALMA detections and the nearest
	{\it Spitzer} YSO.  The full sample is shown by the black outline, while 
	ALMA detections which are highly separated from their nearest SCUBA core
	peak ($>15\arcsec$) are shown in the filled red histogram.
	All sources with {\it Spitzer} separations larger than 15\arcsec\ are included
	in the final bin.}
\label{fig_Spitz_seps}
\end{figure}

In Table~\ref{tab_detects_pt1}, we show the angular separation from each ALMA detection
to the nearest {\it Spitzer} YSO in the catalogue of \citet{Dunham15}.  Figure~\ref{fig_Spitz_seps}
shows these angular separations, which appear to divide into three different regimes:
very small separations ($< 2\arcsec$  between the ALMA detection and a {\it Spitzer}
YSO), moderate separations ($\sim 4\arcsec-7\arcsec$), and larger separations
($> 14\arcsec$).  Some of our ALMA detections also lie significantly separated from 
the closest SCUBA core peak, as shown by the red filled histogram.

If the {\it Spitzer} catalogue of \citet{Dunham15} was complete, we would expect that
all of our ALMA detections separated by more than a few arcsec from the nearest
{\it Spitzer} YSO would be starless.
(We interpret separations of $<$2\arcsec\ as implying completely coincident
ALMA and {\it Spitzer} sources, with the non-zero separation caused by pointing errors,
etc; this is comparable to {\it Spitzer}'s best angular resolution.)
The presence of extended 24~$\mu$m flux and saturated sources
in portions of Ophiuchus, however, likely impacts the completeness of the {\it Spitzer} 
YSO catalogue.
As discussed in \citetalias{Dunham16}, there are several reasons to expect that 
even a complete 
{\it Spitzer} protostellar catalogue does not capture the entire YSO population.
For example, the {\it Spitzer} c2d program \citep{Evans09} missed identifying
low luminosity protostars due to insufficient sensitivity \citep{Dunham08}, and these missing
protostars will affect the classification of their host dense cores.
Indeed, \citet{Schnee10}
detected two protostars with CARMA in their sample of eleven apparently starless cores
in Perseus, both
of which were too faint to have been detected by {\it Spitzer} c2d.
\citet{Stutz13} also find that some Orion protostars above a luminosity of 
$\sim$1~L$_{\odot}$ are too faint at wavelengths of 24~$\mu$m and shorter 
to be identifiable in {\it Spitzer}'s Orion maps. {\it Herschel}, however, is able to 
detect them
clearly at longer wavelengths.
We searched through the literature to look for any evidence that the ALMA detections
separated by more than 3\arcsec\ from a \citet{Dunham15} 
{\it Spitzer} YSO may harbour hidden protostars.

\subsection{Unidentified low luminosity protostars}

\subsubsection{{\it Herschel}}

We first searched for signs of compact protostellar emission 
in {\it Herschel} 
70~$\mu$m and 100~$\mu$m observations of 
Ophiuchus, taken as part of the {\it Herschel} Gould Belt Survey \citep{Andre10},
where the angular resolution is $\sim$5\arcsec.
Most of our ALMA detections not associated with a {\it Spitzer} YSO
appear to have compact detections in the {\it Herschel} data, indicative
of protostars.
The one clear exception is ALMA source 37 
which shows no
signs of a protostellar association.  A further three sources, ALMA sources 
10, 11, and 12
are found in an area 
with complex, extended emission in the {\it Herschel} bands,
making it difficult to ascertain the protostellar
or starless nature of the individual detections.
All four sources without clear protostellar counterparts in {\it Herschel} 
have {\it Spitzer} separations of $>$14\arcsec\ in Figure~\ref{fig_Spitz_seps}.

\subsubsection{Published Catalogues}
We also examined the full, uncut {\it Spitzer} c2d catalogue\footnote{
Available at {\tt http://irsa.ipac.caltech.edu/data/SPITZER/C2D/}.} 
to search for point sources
which were excluded from the final YSO catalogue due to 
confusion from extended emission or other selection criteria.  Most of the {\it Herschel}
protostar detections are also listed in the full {\it Spitzer} c2d
catalogue, with a range of classifications, 
highlighting the challenges of complete YSO identification in Ophiuchus.  
Of special note is that ALMA source 12 has a close positional match with 
{\it Spitzer} source J162630.4-242257, listed as a candidate YSO.  This source is not
present in the YSO candidate catalogue of \citet{Evans09}, implying it failed to 
satisfy one or more additional criterion used to create the final catalogue.
Given the positional match with ALMA source 12, we consider this 
{\it Spitzer} detection to be a likely YSO.

We furthermore searched for associations between the 
the eight ALMA detections located more than
14\arcsec\ from {\it Spitzer} YSOs and
YSO indicators in the literature via 
the SIMBAD database and other published catalogues.

In SIMBAD \citep{Wenger00}, 
we only found three cases of strong evidence for 
positional coincidence with a YSO.
Three of the eight ALMA detections separated by more than 14\arcsec\ from
a {\it Spitzer} YSO are located within 0.6\arcsec\ of an 
X-ray or joint X-ray and radio detection, and are likely to be
YSOs.  ALMA sources 10 and 11
both appear in {\it Chandra} (resolution 0.5\arcsec) and 
VLA (resolution $\ge$ 2.6\arcsec) observations obtained by \citet{Gagne04}
and as such are very strong YSO candidates.  Source 23 was
detected by XMM Newton (resolution 6\arcsec) 
and classified as a candidate star (but not a pre main sequence
star) by \citet{Lin12}.  There is also a T Tau type star located 0.4\arcsec\
from this ALMA detection, which perhaps is tied to the X-ray detection.
Sources 5, 12, 27, and 37 have only tentative associations 
with early infrared-based catalogue objects which, if representing truly young protostellar
sources, we would expect would be easily detectable with {\it Spitzer}.
Source 19 has no match within 14\arcsec\ listed in SIMBAD.

None of the eight ALMA detections 
are listed as sources of molecular
hydrogen outflows identified in \citet{Zhang13}.

Table~\ref{tab_detects_pt1} summarizes all of the strong lines of evidence
of a protostellar association (i.e., {\it Herschel}, {\it Spitzer}, and X-ray plus radio 
catalogues) for each ALMA detection.

\subsubsection{ALMA CO}
If a young and faint protostar is associated with any of the ALMA detections, we
might also see evidence of a protostellar outflow in our ALMA CO data.  
We defer a full analysis of the CO data for future work, but discuss here
results from a quick check of the CO data for the eight ALMA detections located
far from {\it Spitzer} YSOs.
We ran a simple automated version of clean
on the CO observations of the appropriate fields.  We see no evidence for outflows
around any of the positions of the eight ALMA detections.
This lack of detection may not provide a strong constraint on the absence of a protostar, 
however, as without 
compact array or total power observations, it may be difficult to identify outflow emision.

\subsubsection{Summary}
Combining the results from all of these searches we find evidence
that all of the ALMA detections separated by 4\arcsec\ to 7\arcsec\ to the 
nearest \citet{Dunham15} {\it Spitzer} YSO are protostellar.
Nearly all of the eight ALMA detections separated by at least 14\arcsec\ from
the nearest {\it Spitzer} YSO are also protostellar.  Sources 
10, 11, 12, 17, 19, 23, and 27
all have evidence suggesting positional association with a protostar.
Source 37 
is a strong starless core candidate, and  
has a small (3.3\arcsec) separation from the nearest SCUBA core
peak, as expected for a starless core detection.  
These results are summarized in Table~\ref{tab_detects_pt1},
which includes multiple lines of evidence of a protostellar nature
where appropriate.

\subsection{Candidate Starless Core Detections}

\begin{figure}[htb]
\includegraphics[width=3in]{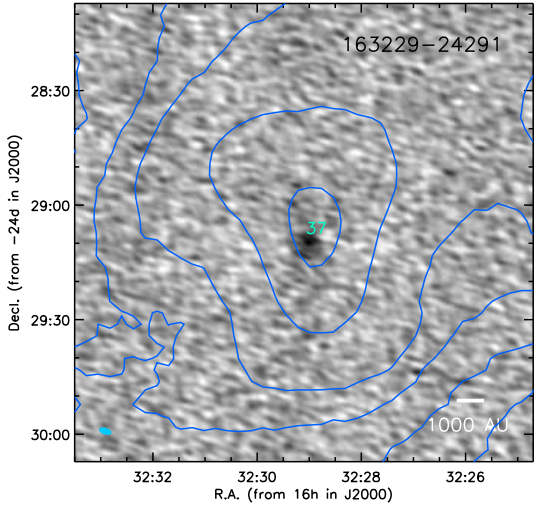}
\caption{
	ALMA single pointing detection of source 37,
	a starless core in field 163229-24291.  
	The greyscale ranges 
	from -0.5~mJy~beam$^{-1}$ to 1~mJy~beam$^{-1}$, and the beam is shown in the bottom 
	left corner.  Blue contours show SCUBA-2 850~$\mu$m emission at levels of 
	0.15~mJy~sq.~arcsec$^{-1}$, 0.5~mJy~sq.~arcsec$^{-1}$, 1~mJy~sq.~arcsec$^{-1}$, 
	1.5~mJy~sq.~arcsec$^{-1}$, 3~mJy~sq.~arcsec$^{-1}$, and 5~mJy~sq.~arcsec$^{-1}$.  
	The horizontal line indicates a length scale of 1000~AU 
	for a distance to Ophiuchus of 140~pc.
	Compact sources of emission detected are labeled with their
	number given in Table~\ref{tab_detects_pt1}.
	This detection coincides with the starless core L1689N
	from \citet{Lis16}, while the closest known protostar is IRAS~16293-2422, which
	lies outside of the area plotted.
	}
\label{fig_single_detects}
\end{figure}

Figure~\ref{fig_single_detects} shows the ALMA image for 
source 37, our best candidate starless core detection, which was 
detected at a level of $4.9~\sigma$.  \citet{Sadavoy10b} identified this starless
core as one of only a handful in Ophiuchus, Taurus, Perseus, and Orion, that appear
to be super-Jeans (having a mass of roughly twice the Jeans mass), and starless,
and suggest it is a strong candidate for collapse.  \citet{Lis16} also identified a
source coincident with our detection using a combination of ALMA Compact Array and CSO
observations, and list it as the starless core L1689N.

\subsection{Extended Emission}
\label{sec_extended}

\begin{figure}[htb]
\includegraphics[width=3.5in]{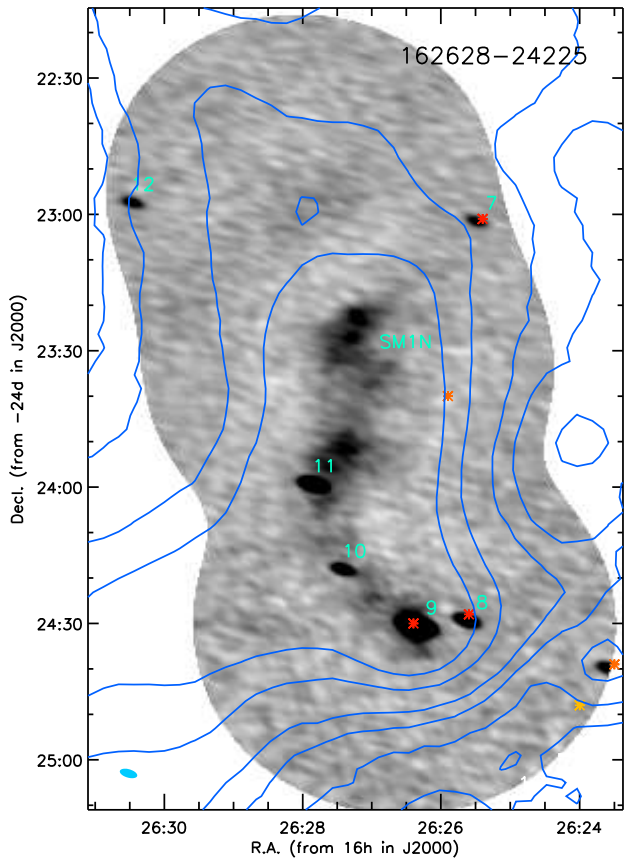}
\caption{
	ALMA mosaic with 
	diffuse extended emission which appears
	somewhat filamentary.  All compact ALMA detections
	in the field are associated with protostellar
	sources.  See Figure~\ref{fig_single_detects} 
	for the plotting conventions adopted.
	In this figure and subsequent ones, the greyscale ranges from
	-1~mJy~beam$^{-1}$ to 2~mJy~beam$^{-1}$.
	Coloured asterisks show {\it Spitzer} YSOs from 
	\citet{Dunham15}, with shading ranging from red for Class~0/I to yellow for Class~II.
	The compact emission in the approximate centre of the mosaic
	(near source 11)
	coincides with a compact source seen in ALMA Band~7 continuum 
	emission by \citet{Friesen14},
	a source known as SM1.  \citet{Friesen14} also detect compact emission in
	the vicinity of the diffuse emission directly east of the SM1N label,
	the compact emission may be obscured in our map by the bright diffuse emission present.
	The southernmost two Class 0/I ALMA detections (sources 8 and 9) 
	are VLA~1623 and VLA~1623W.
	}
\label{fig_mosaic_detects}
\end{figure}

In addition to the compact sources, two of the mosaics show signs of extended
emission.  The more obvious of these is shown in Figure~\ref{fig_mosaic_detects},
where the dense ridge around Oph-SM1 (near source 11) is clearly visible.  
There appears to be three roughly
equally spaced fragments along this ridge, as was also observed by \citet{Nakamura12}
using data from the Submillimeter Array (SMA).  This complex emission structure is
reminiscent of that predicted by the turbulent fragmentation scenario.

\citet{Friesen14} observed H$_2$D$^+$ and continuum emission around 360~GHz and 370~GHz 
with ALMA (Band~7) around SM1N and SM1 (the northern 
and middle of the three fragments visible in
our Figure~\ref{fig_mosaic_detects}), finding emission in both locations.  Their continuum
peak in SM1 matches our detection of source 11.  \citet{Friesen14} do not report
extended continuum emission in either SM1N or SM1, but their observations have
a smaller maximum angular scale ($\sim$6\arcsec), which suggests 
that the emission which we
see would have been filtered out.  \citet{Friesen14} suggest that the compact emission
around SM1 is most likely attributable to a warm accretion disk around a very young
protostar.  Meanwhile, SM1N is more likely to be still
starless, although the presence of a low-luminosity
protostar is also possible.  The extended emission that we detect around SM1N is
therefore a candidate for an additional starless core detection.

\begin{figure}[htb]
\includegraphics[width=4in]{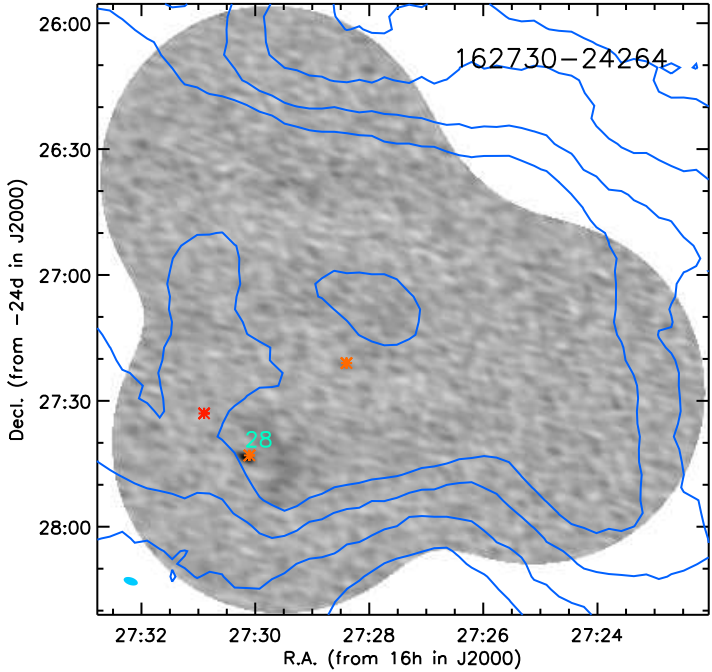}
\caption{ALMA mosaic showing both a compact 
	emission source (source 28) and
	faint extended emission in the southeast area of the map.  The extended
	emission appears to be warm dust which is being heated by the neighbouring
	protostar.
	See Figures~\ref{fig_single_detects} and \ref{fig_mosaic_detects} for the plotting conventions used.
	}
\label{fig_mosaic_extended}
\end{figure}

The second mosaic where faint extended emission is visible is shown
in Figure~\ref{fig_mosaic_extended}.  Both the extended emission and a compact source 
(source 28)
are in close proximity to a Class Flat YSO.  Examination of the {\it Herschel}
data in this field suggests that the extended emission is associated with warm dust
that is being heated by the neighbouring protostar.

\section{Properties of the Detections}

\subsection{SCUBA Core Concentration}
\begin{figure}[htb]
\includegraphics[width=3.5in]{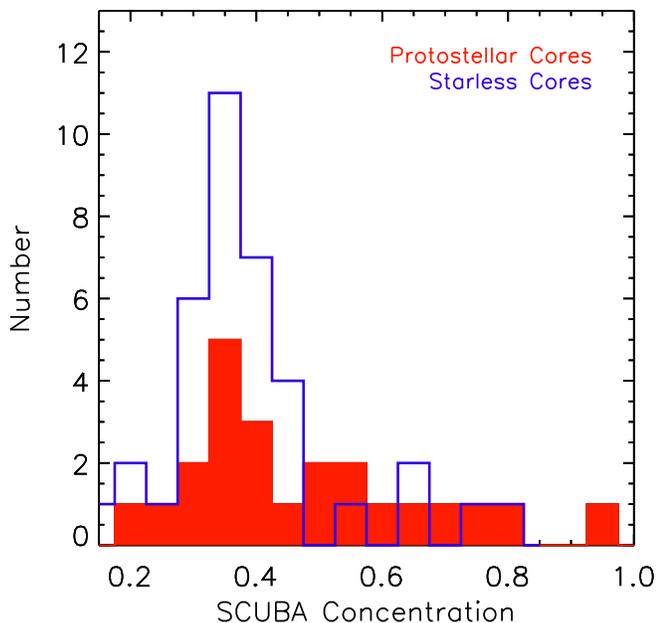}
\caption{
	Distribution of concentrations measured for dense cores within Ophiuchus.
	The dark blue line indicates cores that we classify as starless, while the
	filled red histogram indicates cores that we classify as protostellar.
	}
\label{fig_concs}
\end{figure}

The central concentrations, or measures of the peakiness, of the 
SCUBA dense cores we targeted
were determined by \citet{Jorgensen08}.
Concentration is defined as
\begin{equation}
C = 1 - \frac{1.13 B^2 F_{tot}}{\pi R^2 F_{pk}}
\end{equation}
where $B$ is the beamsize, $F_{tot}$ is the total flux, $R$ is the radius, 
and $F_{pk}$ is the
peak flux \citep{Johnstone00b}. 
A histogram of the SCUBA core concentrations is 
shown in Figure~\ref{fig_concs}.
Although there is a wide 
range of concentrations in both the protostellar and starless core
populations, most of the high concentration cores tend to be protostellar. 
In their analysis of the Perseus molecular cloud, \citet{Jorgensen07} found that
all cores with concentrations above 0.6 were associated with {\it Spitzer}-identified
protostars.  
This same trend, however, did not hold in their similar analysis of 
Ophiuchus, and \citet{Jorgensen08} identified there four 
apparently starless cores with high concentrations.

Source 37, our starless core detection shown in Figure~\ref{fig_single_detects}, 
is found in one of these four
high concentration SCUBA starless cores, with a concentration of 0.74 measured in
\citet{Jorgensen08}.
The remaining three high concentration SCUBA starless cores all lie within the mosaic
shown in Figure~\ref{fig_mosaic_detects}.
The first of these is associated with ALMA source 10, 
a detection where X-ray observations suggest there is a faint hidden protostar.
The second is core 162627-24233, the northernmost core in the 
Figure~\ref{fig_mosaic_detects} mosaic.  Although ALMA did not detect any emission
near the SCUBA core's peak, it did find two sources of compact emission significantly offset
to the west and east of the SCUBA peak, sources 7 and 12 respectively.  
Source 7 is associated
with a \citet{Dunham15} {\it Spitzer} YSO, while source 12
is associated with a {\it Spitzer} YSOc source in the full uncut {\it Spitzer} c2d catalogue
(see Section~3.3).
The final high concentration SCUBA starless core in 
Figure~\ref{fig_mosaic_detects} is not directly associated with any of the 
compact emission sources seen with ALMA, however, it does contain the extended emission
seen around SM1N discussed earlier.

Even if these high concentrations do
not imply the presence of hidden faint protostars, they are expected to be 
indicative of more-evolved cores, with the presence of a more concentrated peak
of emission expected to occur near the time when a starless core begins to collapse
to form a protostar.  Our observations reinforce this picture.
None of the SCUBA starless cores with concentrations below 0.6 
were detected with ALMA, while all four of the starless cores with concentrations above 0.6
have at least one associated detection, and two of these four `starless' cores 
(associated with ALMA sources 10 and 12) may already harbour faint protostars.

\subsection{ALMA Peak Fluxes}
\begin{figure}[htb]
\includegraphics[width=3.5in]{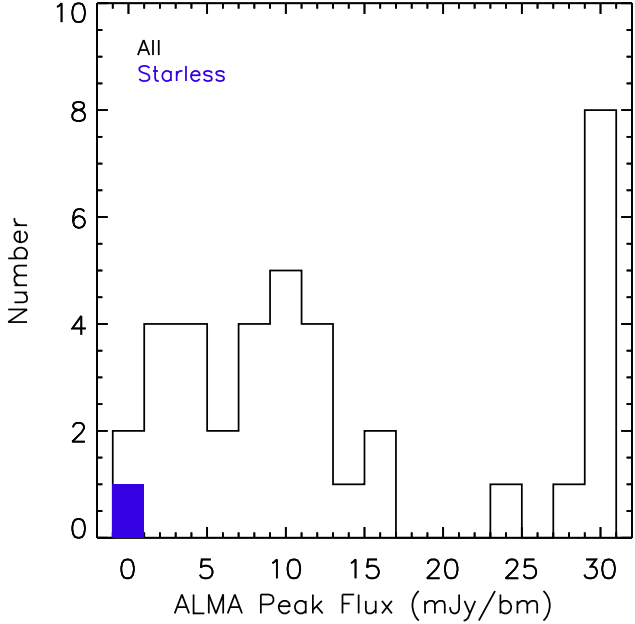}
\caption{
	Distribution of peak fluxes of ALMA detections.  Sources with peak fluxes higher
	than the maximum bin size plotted are included in the final bin.
	The black empty histogram shows the 
	entire 38 ALMA detections, while the filled blue histogram shows 
	source 37, the starless core detection.
	}
\label{fig_alma_pkflux}
\end{figure}

In Figure~\ref{fig_alma_pkflux}, we show the distribution of peak
fluxes measured for the ALMA detections.  The filled blue histogram shows
source 37 which appears to be starless.  Notably, source 37 lies
in the lowest peak flux bin.
Since starless cores are expected to pass through
higher density phases more rapidly (see discussion in the next section), it
is reassuring to see source 37 lying at the low end of the distribution.

We note that if the diffuse emission we see around SM1N were included as a source in 
Figure~\ref{fig_alma_pkflux}, it would lie near source 37 with a peak flux around 
2~mJy~beam$^{-1}$.

\section{Substructure from Turbulent Fragmentation}
We now examine the number of starless cores that 
we would expect to detect under the
turbulent fragmentation picture.
For this analysis, we take a similar approach to that outlined in
\citetalias{Dunham16}.  The premise is that numerical simulations of
the evolution of dense cores 
may provide a better 
guide to the detectability of substructure within dense
cores than would a simple
model such as a collapsing Bonnor Ebert sphere model.  \citetalias{Dunham16}
also analyze the detectability of a collapsing Bonnor Ebert sphere,
and find that it spends little time in a highly dense and concentrated state. 
Over an ensemble of cores at different ages, a smaller fraction 
of Bonnor Ebert spheres should be detectable than substructure generated
in the numerical simulations including turbulence and / or magnetic fields.

\citetalias{Dunham16} analyze simulations of isolated collapsing 
prestellar cores performed using the {\sc Orion} adaptive mesh refinement code
\citep{Li12}. 
The calculations include turbulence, an initially uniform vertical magnetic field and 
self-gravity.  Further evolution of cores with these initial conditions finds that secondary 
fragmentation, leading to wide-binary formation, is common \citep{Offner16}. 
\citetalias{Dunham16}, however, only perform the analysis until the first star 
(sink particle) forms, at which point collapse has produced substructure but not 
multiplicity.
Using synthetic observations of these simulations,
\citetalias{Dunham16} demonstrate that for their ALMA Cycle 1 setup they expect
to detect cores with central densities above $8.9 \times 10^7$~cm$^{-3}$.  This
value is derived based on synthetic observations of a dense core with a 
total mass of 0.4~\Msol; a higher mass core (4~\Msol) would be
detectable when its central density exceeds $3.3 \times 10^7$~cm$^{-3}$.
We note the simulation core masses are upper limits compared to the observations, since they 
do not account for observational detection limits or consider how SCUBA cores are defined. 

The detection threshold is also a function of both the sensitivity of the array
and the spatial filtering implied by the antenna configuration.
We therefore re-run the synthetic observations presented in \citetalias{Dunham16}
to account for the antenna configuration used for our Ophiuchus observations, in addition
to scaling the simulations to the slightly closer distance of 140~pc for Ophiuchus
(compared with 150~pc for Chamaeleon).  \citetalias{Dunham16} used a column density-temperature
relationship derived from Bonnor Ebert sphere models estimate the total observable flux
implied by the simulations.  The core interiors that we observe are heavily shielded from 
the radiation field impinging on the molecular cloud exterior.  The temperatures
on the sub-core scales of interest should therefore be similar in both Chamaeleon and Ophiuchus, 
and we apply the same relationship as in \citetalias{Dunham16}.

In Ophiuchus, adjusting the \citet{Jorgensen08} core catalogue values to our
assumed distance of 140~pc, we find a median core mass of 0.55~\Msol.
Hence, as in \citetalias{Dunham16}, the 0.4~\Msol\ core is the most applicable
simulation to use for the synthetic observations.

\begin{figure}[htb]
\includegraphics[width=6in]{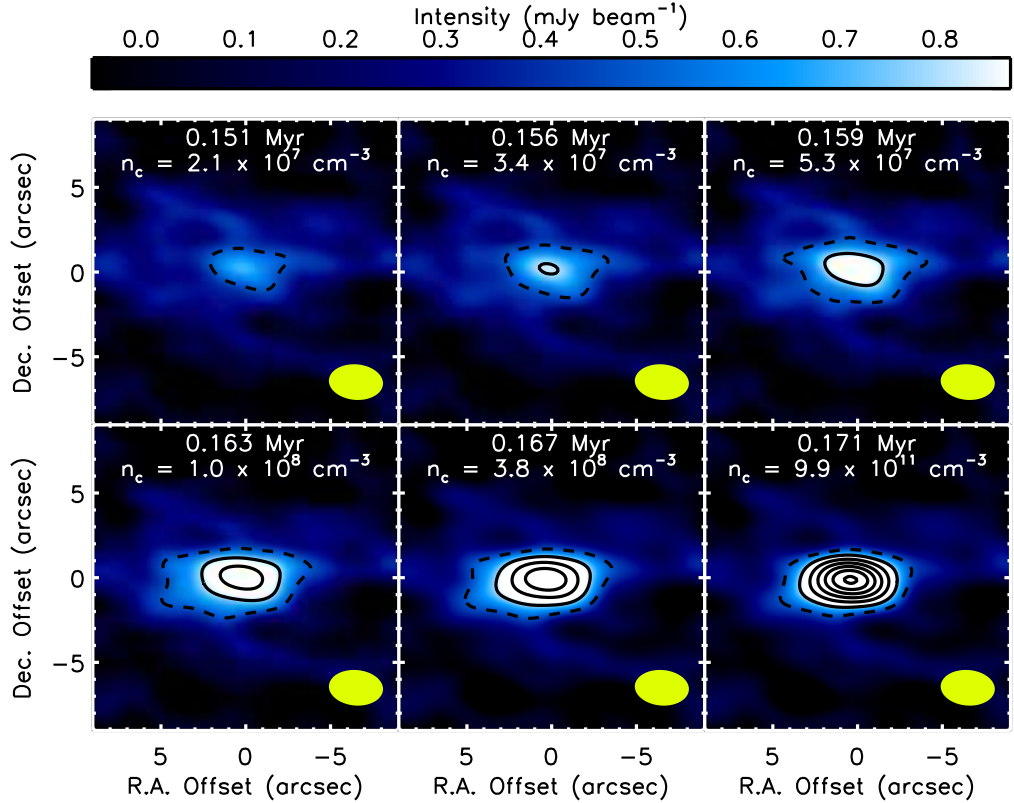}
\caption{Synthetic ALMA 107~GHz observations of the evolution of a simulated 0.4~\Msol\ core 
	at six time steps.  The dashed contour represents the 3$\sigma$ level, while the
	solid contours show levels of 5$\sigma$ and above, with separations of 2$\sigma$
	between each level.  In our observations 1$\sigma \sim 0.16$~mJy~beam$^{-1}$.  The
	beam is indicated by the solid yellow ellipse.  A robust (5$\sigma$) detection of
	the core occurs after 0.156~Myr, when the central density of the core reaches
	$3.4 \times 10^7$~cm$^{-3}$.}
\label{fig_synth}
\end{figure}

Figure~\ref{fig_synth} shows our synthetic ALMA observations of the 0.4~\Msol\ core simulation,
adopting a distance of 140~pc and the antenna configuration used for our real observations.
Similar to the synthetic observations generated by \citetalias{Dunham16},
the synthetic observations were generated using the CASA tasks
{\sc simobserve} and {\sc simanalyze}.  We assigned a position of R.A. = 16:30:00 and
decl. = -24:30:00, to the simulations, representing the middle of the area observed in Ophiuchus.  
We assumed 1~s integration times
and included thermal noise from the atmosphere, for a total integration time of
60~s, with an effective mean frequency of 107~GHz and
bandwidth of 6~GHz.  We cleaned to a threshold of 0.30~mJy~beam$^{-1}$, 
i.e., approximately 2$\sigma$, using non-interactive cleaning with no clean mask set a priori.
The synthetic observations do not account for the flagging applied to the real observations,
which was extensive in our data, nor the exact elevation or weather conditions that we observed 
under.  Thus, the synthetic observations have a somewhat lower noise level than our real 
observations do.  We only applied clean and searched for detections in terms
of our typical observed noise level to account for this difference in data quality.

As can be seen in Figure~\ref{fig_synth}, the central core peak is only detected in the 
simulations at a 5$\sigma$ level after 0.156~Myr, when the central density of the core
reaches $3.7\times 10^7$~cm$^{-3}$.  This is an earlier and less dense stage of the core's
evolution than \citetalias{Dunham16} found for the first core detection in Chamaeleon-like
cores.  The primary driver for this difference arises from our observations having more antennae
at shorter baselines, improving the sensitivity to larger-scale structure.

In order to predict the number of cores detected based on the observational set-up,
\citetalias{Dunham16} then assumed that the lifetime of a dense core varies with
its freefall time \citep[see, e.g.,][]{Jessop00}.
Under the important
assumption that star formation is continuous in the region, with the dense core
lifetime at different densities described by the freefall time, 
\citetalias{Dunham16} express the number of ALMA detections expected as
\begin{equation}
Detections > \frac{2}{3} \times N_{total} \times \big(\frac{n_{Detectable}}{n_{Limit}}\big)^{-0.5}
\label{eqn_dets}
\end{equation}
where $N_{total}$ is the number of starless cores observed, 
$n_{Detectable}$ is the
density at which ALMA is able to detect the starless cores, and $n_{Limit}$ is the
observed lower limit of the mean core densities as observed at single-dish resolution.
The factor of two thirds arises from the fact that \citetalias{Dunham16}
note that in the simulations, detections are much more difficult for cores viewed along
the local magnetic field axis.
In the distance-adjusted \citet{Jorgensen08} catalogue, the minimum core density 
to apply in Equation 4 is 
$2.5 \times 10^5$~cm$^{-3}$,
for our adopted temperature and distance, while the mean core density is 
$1.5 \times 10^6$~cm$^{-3}$.

For Ophiuchus, we have $N_{total}=$ 37 starless cores\footnote{The astute
reader will note that in Section~3.3, we found seven ALMA detections which had
large separations from all YSOs in \citet{Dunham15} but were classified as
protostellar rather than starless based
on comparison with ancillary data.  For various reasons, including the large offset
of many of these detections from the nominal dense core centre, and detections which
share a common dense core field, re-classifying seven of the ALMA detections as
protostellar does not imply that there are an equivalent seven SCUBA starless core targets
which should be re-classified as protostellar.  Nonetheless, even if we did assume only
30 starless cores, the resulting detection rate from Equation~4 would still be two.}
with a minimum density of
$n_{Limit} = 2.5 \times 10^5$~cm$^{-3}$, which we expect are detectable with our
ALMA setup down to a density of $n_{Detectable} = 3.7 \times 10^7$~cm$^{-3}$.
Based on equation~\ref{eqn_dets}, we would therefore predict a minimum 
of 2.1 (two) starless cores to have detectable substructure.  
\citetalias{Dunham16} additionally demonstrated the much greater
difficulty of detecting Bonnor Ebert spheres compared with the 
simulated starless cores, when both were observed under identical conditions.
\citetalias{Dunham16} found that a Bonnor Ebert sphere was not detectable until
it had a central density more than 100 times larger than the minimum density
threshold for detecting the simulated cores, implying a detection rate more than
ten times lower for the Bonnor Ebert sphere model.

In our observed sample, we have one detection which has strong evidence suggesting
that it is not associated with a protostellar core.
We also have one detection of extended emission around SM1N which is
tentatively unassociated with any YSOs. 
These detections suggest that the cores are not well-described
by the simple Bonnor Ebert sphere model, as under that scenario, no detections are expected. 
Including both sources as starless core detections, we satisfy Equation~4.  
Considering simple Poisson statistics, the chance of having one or two detections
is equally likely (27\% in both cases), so even if we only consider one detection,
we are not inconsistent with Equation~4.

\subsection{Comparison to Chamaeleon~I}

In their search for substructure within starless cores in Chamaeleon~I, \citetalias{Dunham16}
had no detections, despite expecting at least two based on the arguments
discussed in the previous section.
The \citetalias{Dunham16} null result had reasonably high statistical significance, 
and they discussed three possible reasons for the disagreement.  
The first, which they argued was the most likely, is
that star formation is not continuous in Chamaeleon~I, i.e., the present population of
starless cores is 
not likely to collapse and form protostars in the near future, and
instead may be even 
on their way to dispersing.  
Simple virial estimates of cores in Chamaeleon~I suggest that most of the starless
cores are stable against gravitational collapse, which lends credence to this argument.
The remaining two possible reasons 
for disagreement have more global
implications, and would be also expected to manifest themselves in observations of
starless cores in other molecular clouds.  The second is that the
assumption that the core lifetime is proportional to the freefall time is not correct,
while the third is that the numerical simulations used for the comparisons
are not applicable.

If either of the latter two more global explanations put forward by 
\citetalias{Dunham16} are
true, we could reasonably expect a similar outcome in Ophiuchus, i.e., fewer detections
of substructure than the lower limit that Equation~\ref{eqn_dets} predicts.  
Unless both our one strong and one tentative non-protostellar 
detections are actually
attributable to mis-classified protostellar sources, 
we have identified 
as many starless core substructures as expected. 
This result lends credence to the hypothesis proposed in \citetalias{Dunham16} that the
dense cores in Chamaeleon~I are in an unusual state of evolution which has 
contributed to the lack of detections in that cloud.

Properties of the ambient molecular cloud environment may also play a role,
and perhaps are tied to the difference between Chamaeleon~I and Ophiuchus,
and could imply that the simulations also poorly reflect core evolution
in a Chamaeleon~I-type environment.  
Comparing core detection statistics in different regions implicitly assumes that
cores have similar structures across clouds, which may not be entirely correct.
In single-dish observations, for example, dense cores in Ophiuchus tend to
appear much more compact and clustered than in other nearby clouds.
Dense cores in Ophiuchus show sizes of 14\arcsec\ to 40\arcsec\,
with the majority having sizes under 30\arcsec\ \citep{Pattle15}, 
while dense cores in Chamaeleon~I have sizes ranging from 20\arcsec\ to 140\arcsec\
with the majority having sizes well above 30\arcsec\
\citep{Belloche11}. (Recall that the two clouds have similar distances, of
140~pc and 150~pc respectively.)
Cores which are typically smaller on single-dish 
scales could also have a more compact configuration of mass on smaller scales, which
would then increase their central densities and
likelihood of detection with ALMA.  As seen with our 
Ophiuchus observations, it is also possible that typically smaller and more clustered
cores may live in a generally more complex environment where dense substructures
can form outside of the most obvious density peaks seen in single-dish observations.

\subsection{Comparison to Orion}
\citet{Kainulainen16} also recently examined ALMA continuum
observations of the northern portion of the integral shaped filament in Orion.
They detected 43 compact sources across a mosaic of the filament, including 18 
that do not appear to be associated with a protostar.  Given the high degree of 
nebulosity within Orion, however,
it is difficult to be completely certain of a starless classification.
The number of expected detections also cannot be easily quantified for this study,
as the number and properties of starless cores within the mosaic observed were not
quantified prior to the ALMA observations.  With an appropriate starless core
catalogue and synthetic observations of simulations tuned to the Orion data,
it would be interesting to see how the \citet{Kainulainen16} results compare
to Ophiuchus and Chamaeleon~I.

These ALMA results highlight the fact that starless core populations 
need to be observed across multiple molecular clouds, and preferably in
a uniform manner.  This would increase the detection statistics, allowing for more
stringent tests of models, and also
help to disentangle region-specific behaviour that may be occurring.

\section{Conclusion}
\label{sec_conc}
We present ALMA Cycle 2 Band 3 (3~mm) observations of 60 dense cores in the Ophiuchus
molecular cloud, which were previously identified with SCUBA at the JCMT.
We detect 38 compact emission structures with ALMA, including at least
one detection which has strong signs of being starless.  An additional 
source of diffuse emission, Oph-SM1N may also be starless.  Most of the known protostars
falling within our surveyed area were detected, as discussed in more detail in 
Appendix A.3.

Our ALMA observations should be easily able to detect very low luminosity
protostars or first hydrostatic cores that were undetectable with {\it Spitzer}.
Indeed, we initially identified eight ALMA detections which had no counterparts in 
the \citet{Dunham15} {\it Spitzer} YSO catalogue.  Seven
of these detections showed strong evidence
of a protostellar nature using ancillary data, 
leaving just a single strong candidate starless detection.
We additionally identify extended emission around Oph SM1N, which we
classify as a candidate starless core detection.

One or both 
of our detections of starless candidates may reveal the substructure 
present within
starless cores near the end of their evolution towards protostars.  Following comparisons
with synthetic observations of numerical simulations and simple analytic arguments 
similar to those
presented in \citet{Dunham16}, the number of starless core detections predicted
($>2$) is somewhat consistent with the number we actually detect 
(one strong candidate and one additional candidate).
Both of 
our candidate starless detections are associated with high concentration
SCUBA cores and
are located near a SCUBA core peak, with low ALMA peak flux, as expected for true starless core
detections.
If the starless cores in Ophiuchus had instead been well-described by a smooth 
Bonnor Ebert sphere model, we would have expected no detections.
In Chamaeleon~I,
\citet{Dunham16} detected no starless cores, despite an expectation of detecting at least two
based on a comparison with numerical simulations.  
We argue that the most likely explanation for the difference between Ophiuchus and
Chamaeleon~I cores is one put forward by
\citet{Dunham16}, namely, that the starless cores presently in Chamaeleon~I are not
evolving toward a protostellar phase, and currently do not appear to be gravitationally
unstable.
Observations of additional starless core populations are necessary to confirm whether 
or not
this picture of starless core evolution is correct.  
As ALMA gains further antennas,
new observations should additionally become more sensitive
to multiple turbulence-generated fragments \citep[e.g.,][]{Offner12} if they
exist, informing models of stellar multiplicity.

\appendix
\section{Protostellar Detections}

Here, we show our ALMA protostellar detections, moving through the different 
\citet{Dunham15} {\it Spitzer} YSO separation regimes in Figure~\ref{fig_Spitz_seps} in
turn.
We emphasize that although some of these ALMA detections have large separations
from the nearest {\it Spitzer} YSO in the \citet{Dunham15} catalogue, all show signs of
a protostellar nature in ancillary data.

\subsection{Large {\it Spitzer} Separations}
There are eight ALMA sources in Figure~\ref{fig_Spitz_seps} which have
{\it Spitzer} YSO separations of $>$14\arcsec.  One of these 
was presented in Section~3.4, source 37.
A further three of the eight
sources, source 10, 11, and 12 appear in Figure~\ref{fig_mosaic_detects}.

\begin{figure}[htb]
\begin{tabular}{cc}
\includegraphics[width=3in]{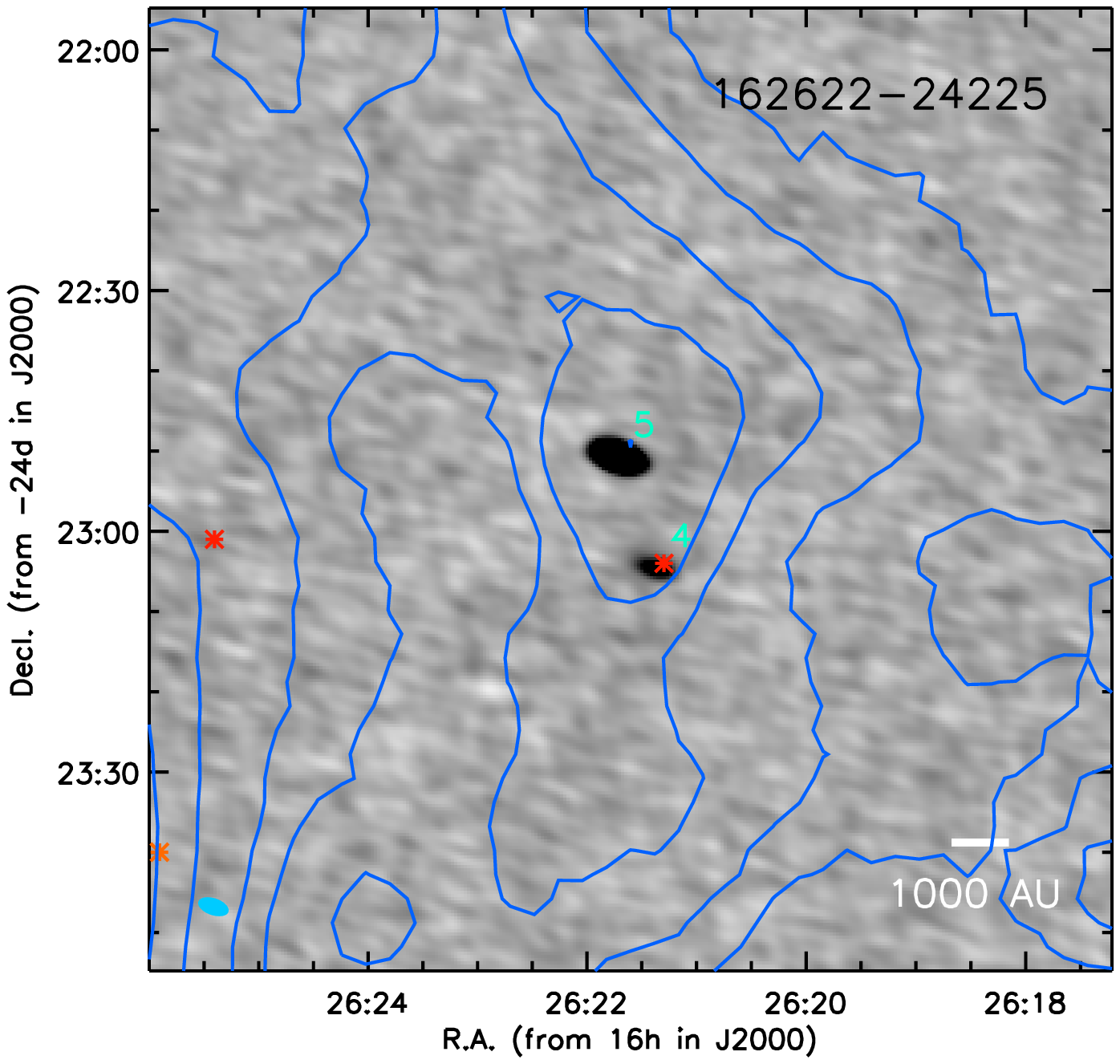} &
\includegraphics[width=3in]{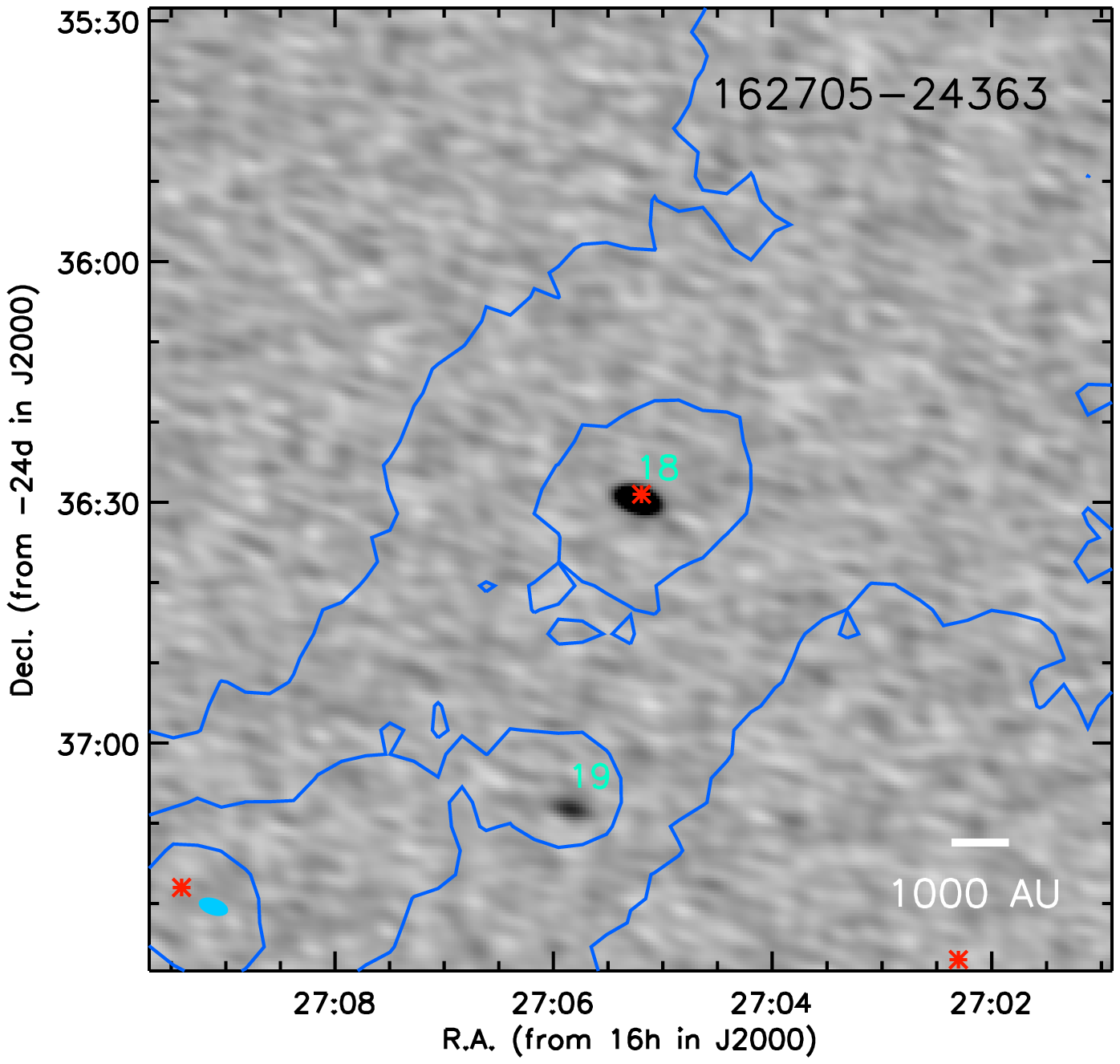} \\
\includegraphics[width=3in]{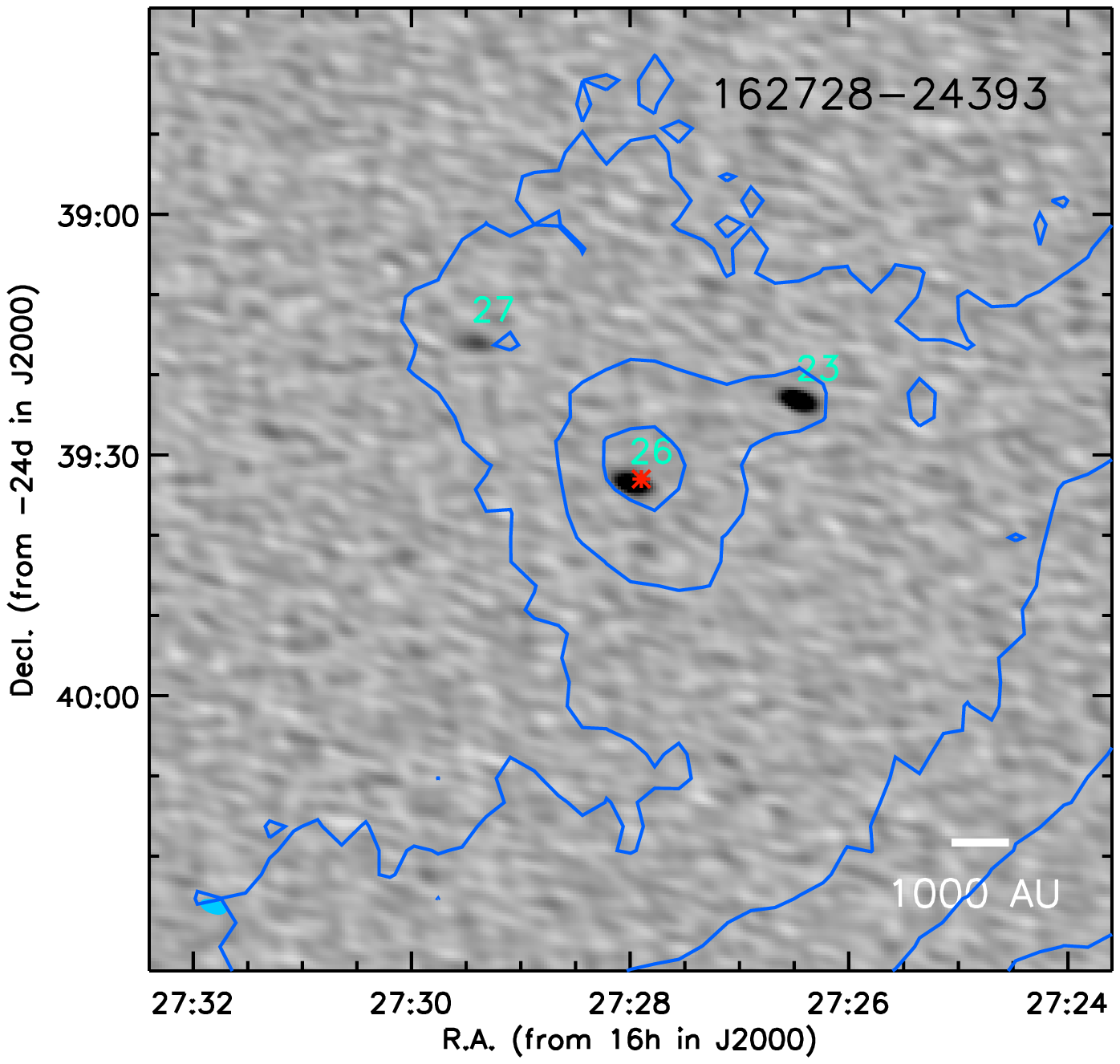} & 
\\
\end{tabular}
\caption{
	ALMA detections of sources separated by at least 14\arcsec\ from
	the nearest {\it Spitzer} YSO, but showing other signs of a protostellar nature.
	See Figures~\ref{fig_single_detects} and and \ref{fig_mosaic_detects}
	for the plotting conventions adopted.
	These sources are source 5, 19, 23, and 27.  
	Other close protostellar associations in the field
	are also labelled.
}
\label{fig_offs_proto_detects}
\end{figure}

Figure~\ref{fig_offs_proto_detects} shows the remaining four ALMA sources,
sources 5, 19, 23, and 27, which lie in three separate fields.
As can be seen in the figure, three of the
four ALMA detections in Figure~\ref{fig_offs_proto_detects} 
appear to be separated from both the SCUBA core peak and a
central {\it Spitzer} YSO.  There are subtle indications for all three ALMA detections
that they lie on local SCUBA-2 flux maxima which were not prominent enough to
be identified as distinct cores in \citet{Jorgensen08}\footnote{We also checked
the \citet{Pattle15} SCUBA-2-based core catalogue, but did not find close positional 
associations between any of these detections and additional cores identified 
by \citet{Pattle15}.}.  

\subsection{Moderate {\it Spitzer} Separations}
\begin{figure}[htb]
\begin{tabular}{cc}
\includegraphics[width=2.6in]{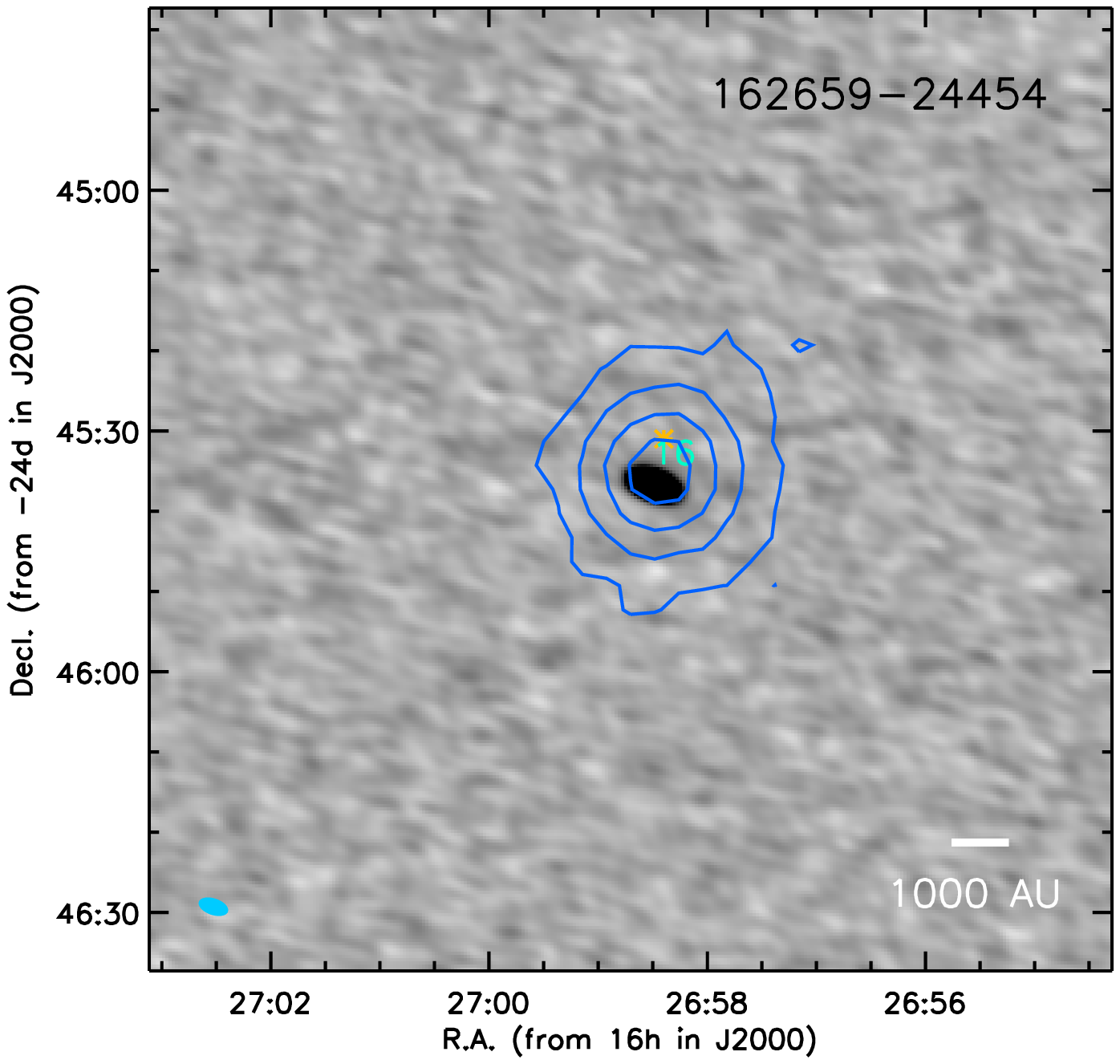} &
\includegraphics[width=2.6in]{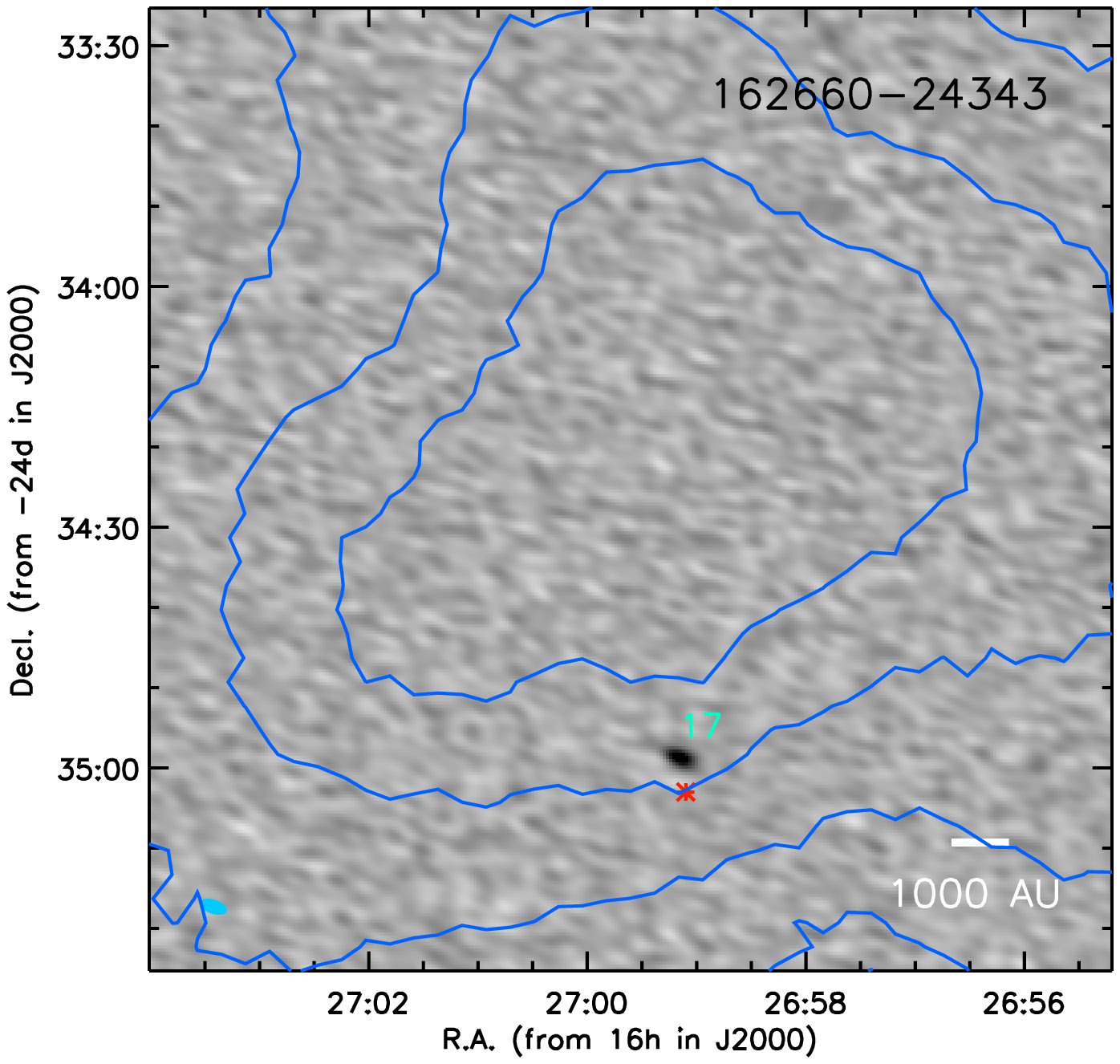} \\
\includegraphics[width=2.6in]{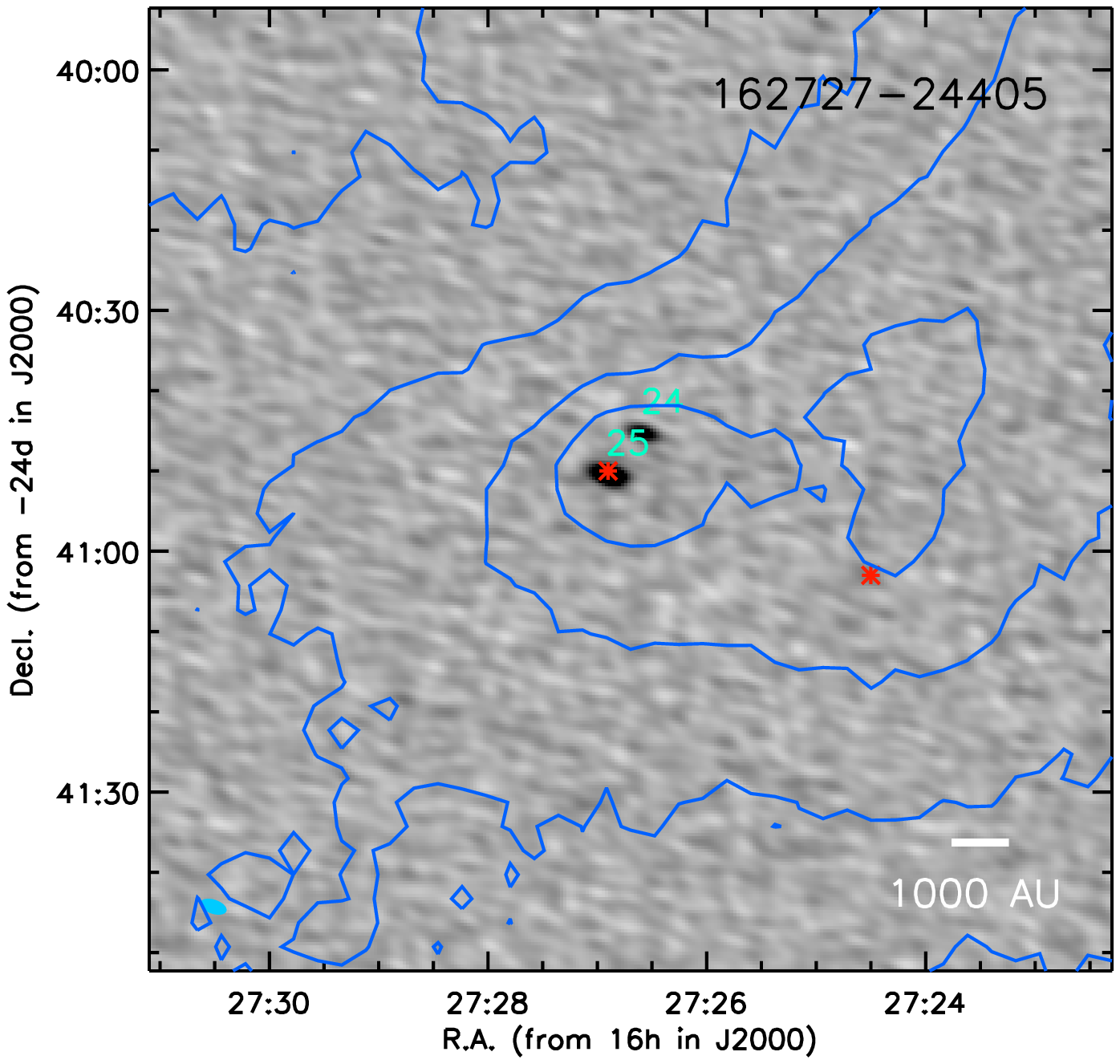} &
\includegraphics[width=2.6in]{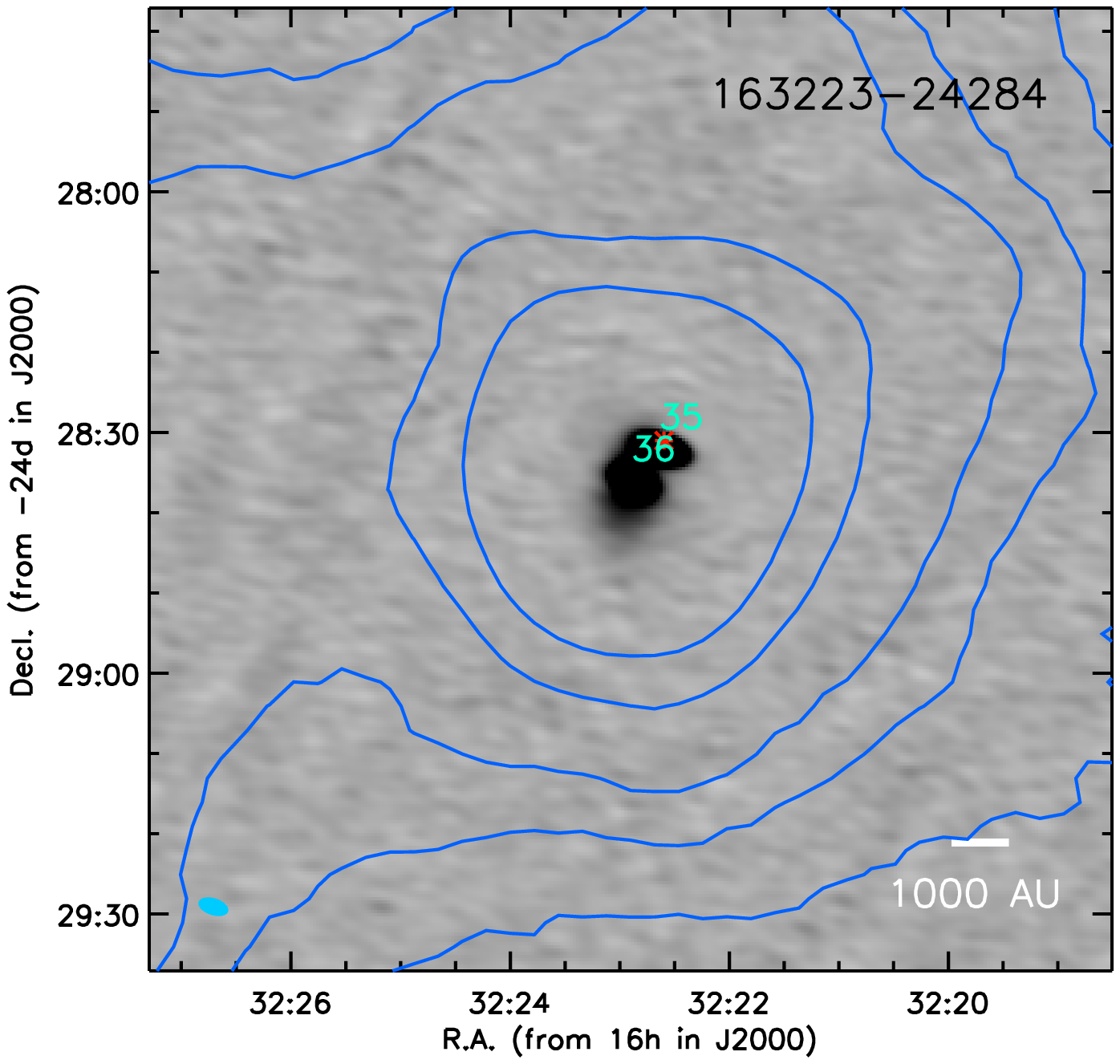} \\
\end{tabular}
\caption{
	ALMA detections where there is a moderate separation between the detection
	and the closest {\it Spitzer} YSO, but there are other signs of a protostellar nature.
	Sources 16, 17, 24, and 36 have moderate {\it Spitzer} separations.
	See Figures~\ref{fig_single_detects} and \ref{fig_mosaic_detects}
	for the plotting conventions adopted.  
	For the field
        163223-24284 only, the range for the greyscale image is -10~mJy~beam$^{-1}$ 
	to 20~mJy~beam$^{-1}$.
	In the top left panel,
	{\it Herschel} shows a a single protostar lying coincident with ALMA source 16,
	suggesting that there is a positional error for the {\it Spitzer} YSO.
	Sources 35 and 36 
	correspond to the binary protostar IRAS~16293-2422; both members of the 
	binary are identified as protostellar in the SIMBAD database.
	}
\label{fig_semioffs_protos1}
\end{figure}

\begin{figure}[htb]
\includegraphics[width=4in]{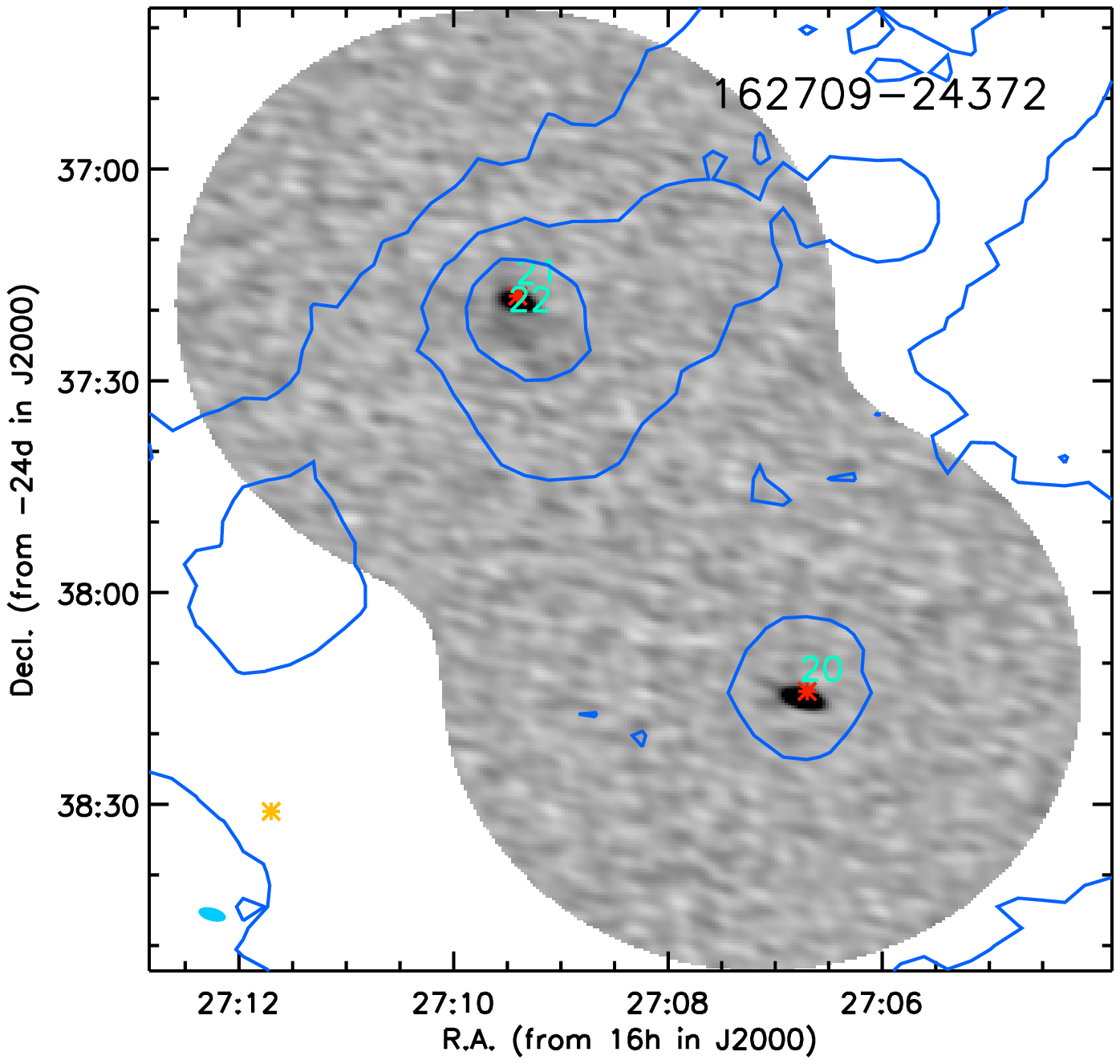}
\caption{
	Mosaic containing an ALMA detection where there is a moderate separation 
	between the detection
	and the closest {\it Spitzer} YSO, but there are other signs of a protostellar nature.
	See Figures~\ref{fig_single_detects} and \ref{fig_mosaic_detects}
	for the plotting conventions adopted.  
	Source 22 is a marginal ($<5~\sigma$) detection, which has a moderate
	{\it Spitzer} separation.  It lies directly south of the bright source 21. 
}
\label{fig_semioffs_protos2}
\end{figure}

In Figure~\ref{fig_semioffs_protos1} and \ref{fig_semioffs_protos2}, 
we show the five ALMA detections which
have moderate separations (4\arcsec\ to 7\arcsec) from the nearest {\it Spitzer} YSO,
for which {\it Herschel} observations revealed protostellar counterparts.


\subsection{Small {\it Spitzer} Separations}
\begin{figure}[htbp]
\begin{tabular}{cc}
\includegraphics[width=2.9in]{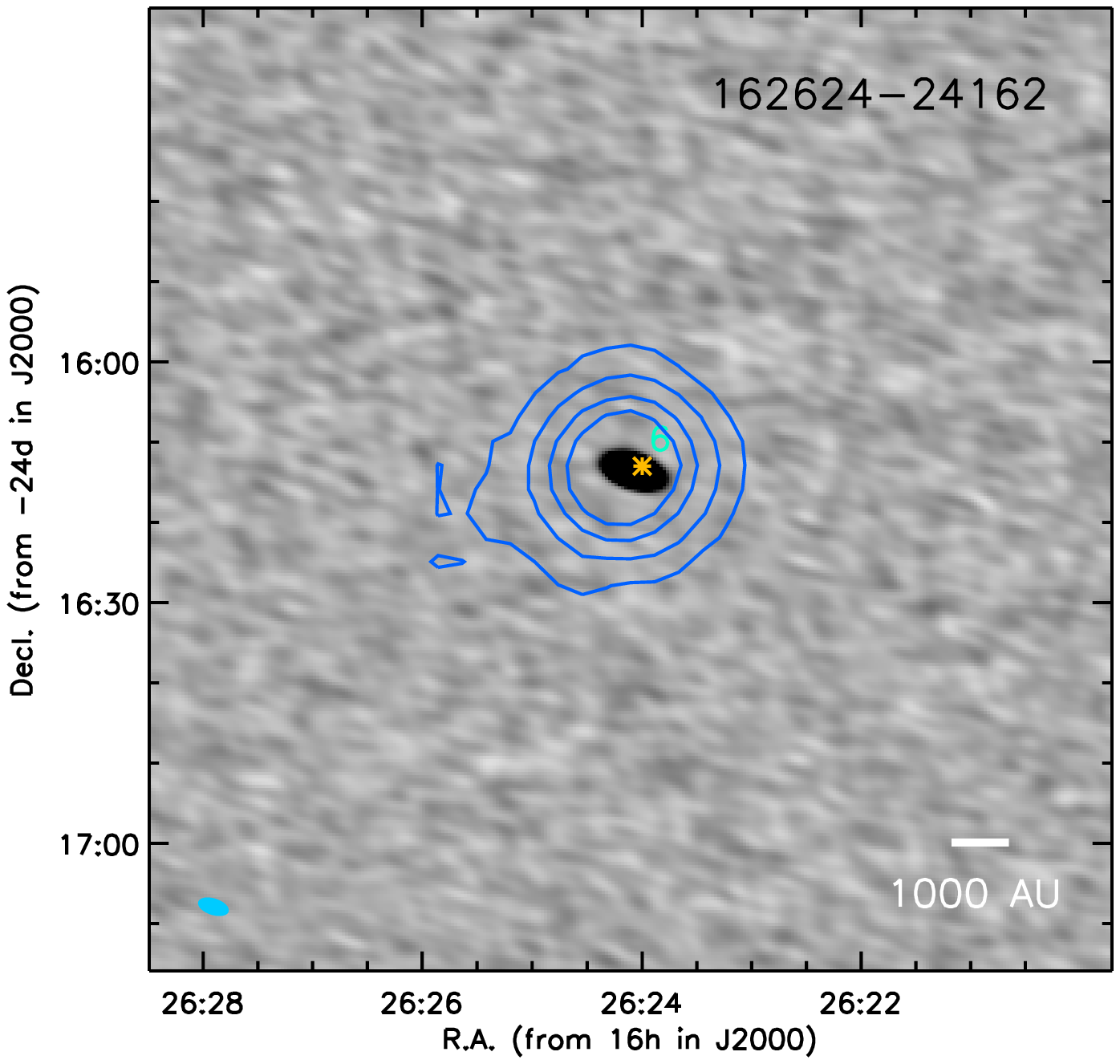} &
\includegraphics[width=2.9in]{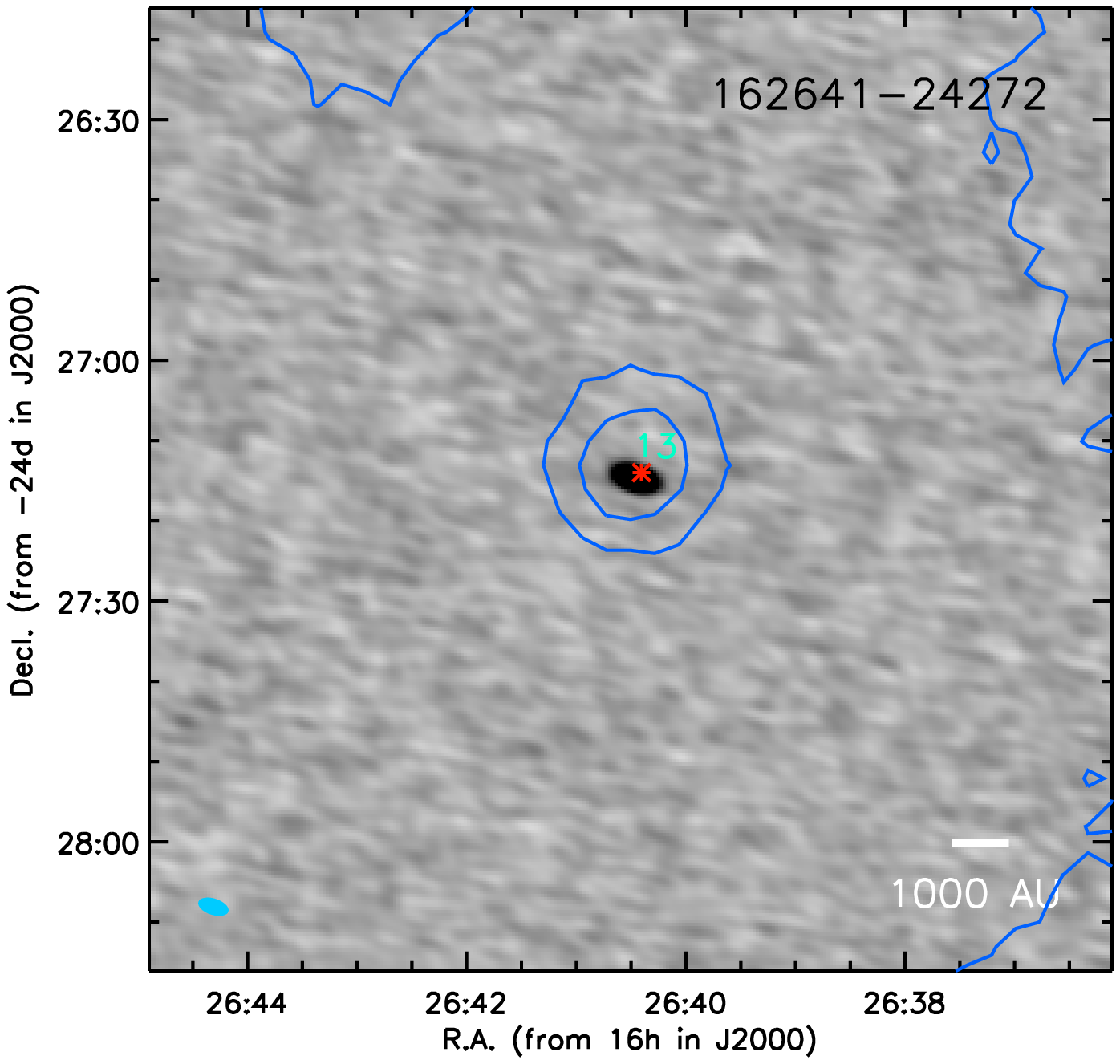} \\
\includegraphics[width=2.9in]{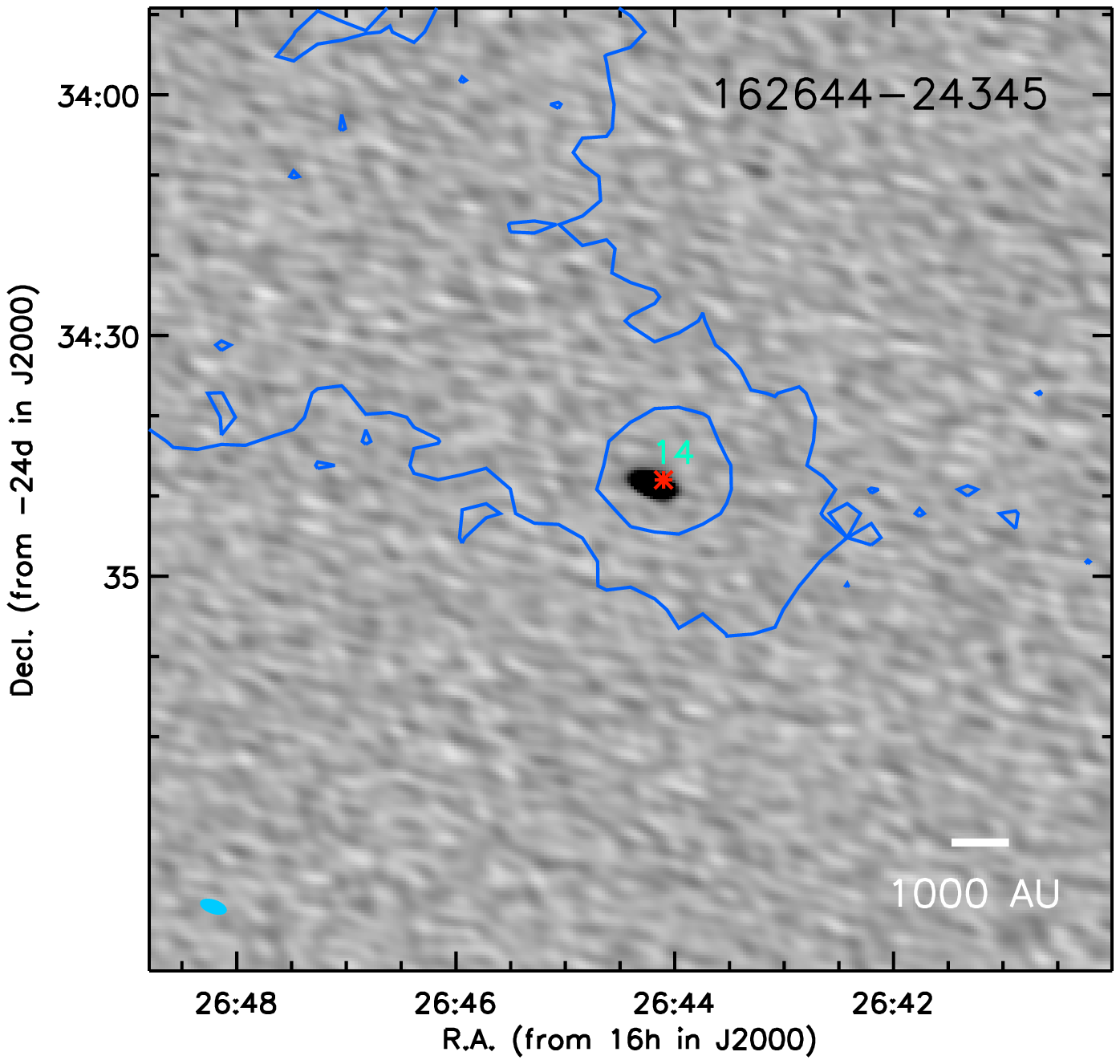} &
\includegraphics[width=2.9in]{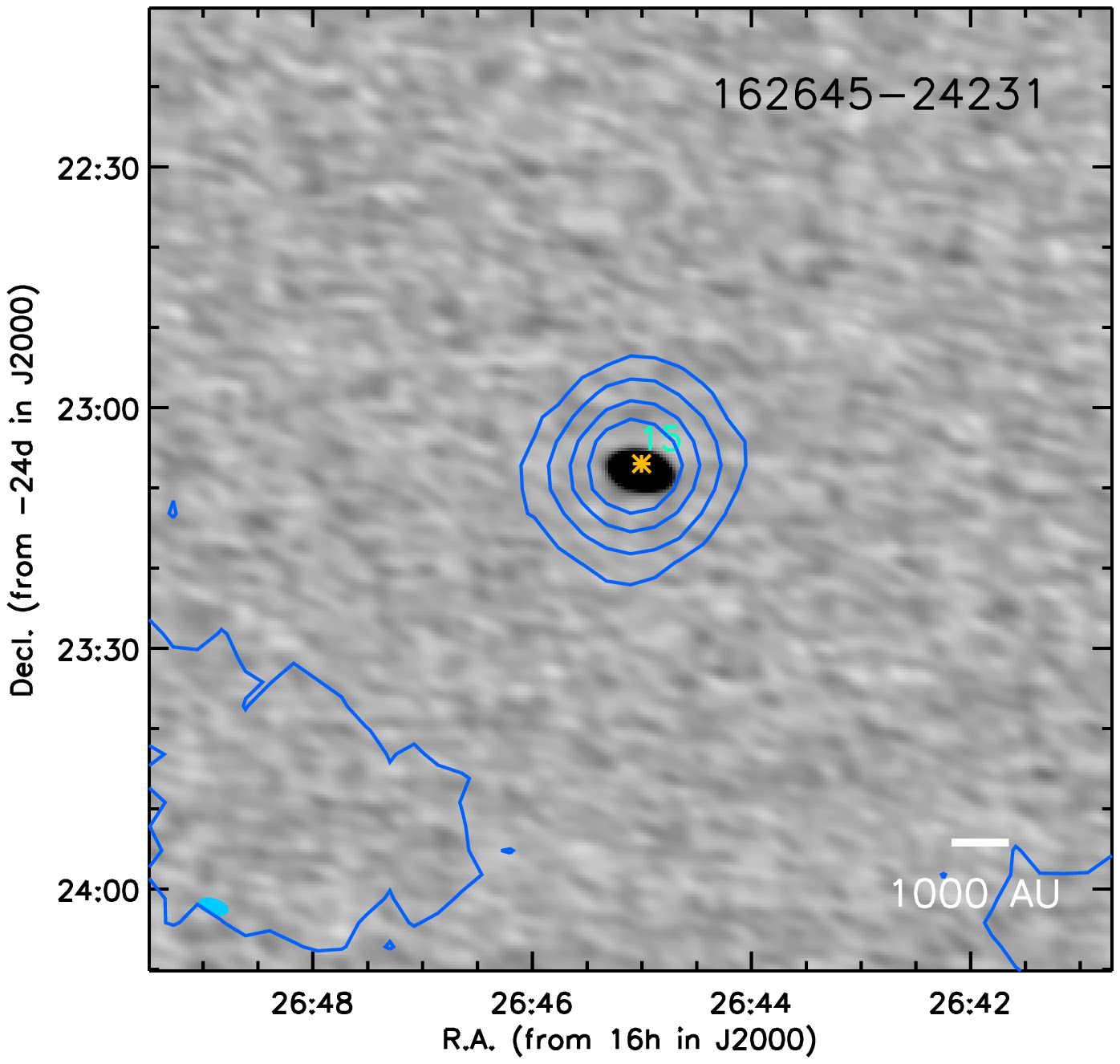} \\
\includegraphics[width=2.9in]{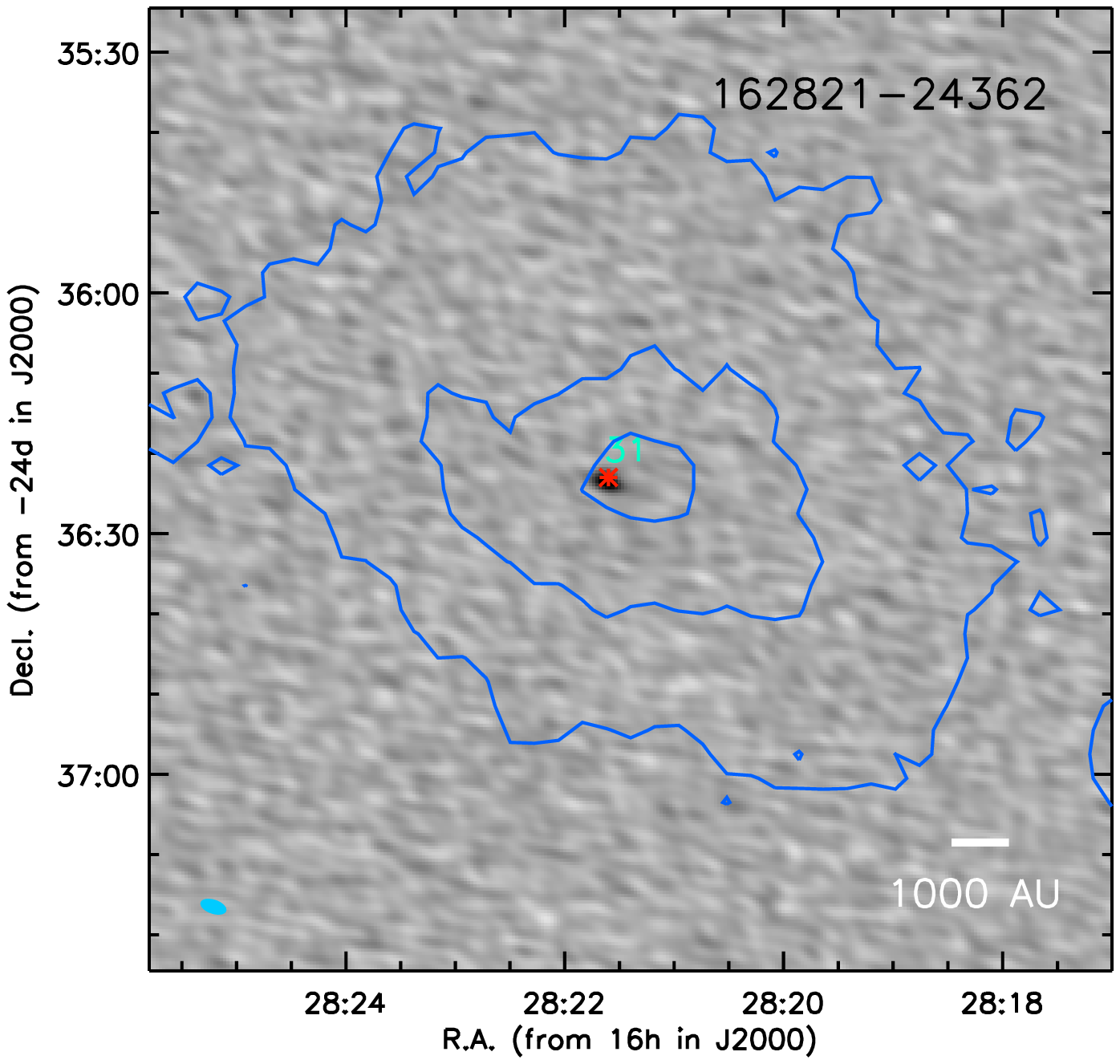} &
\includegraphics[width=2.9in]{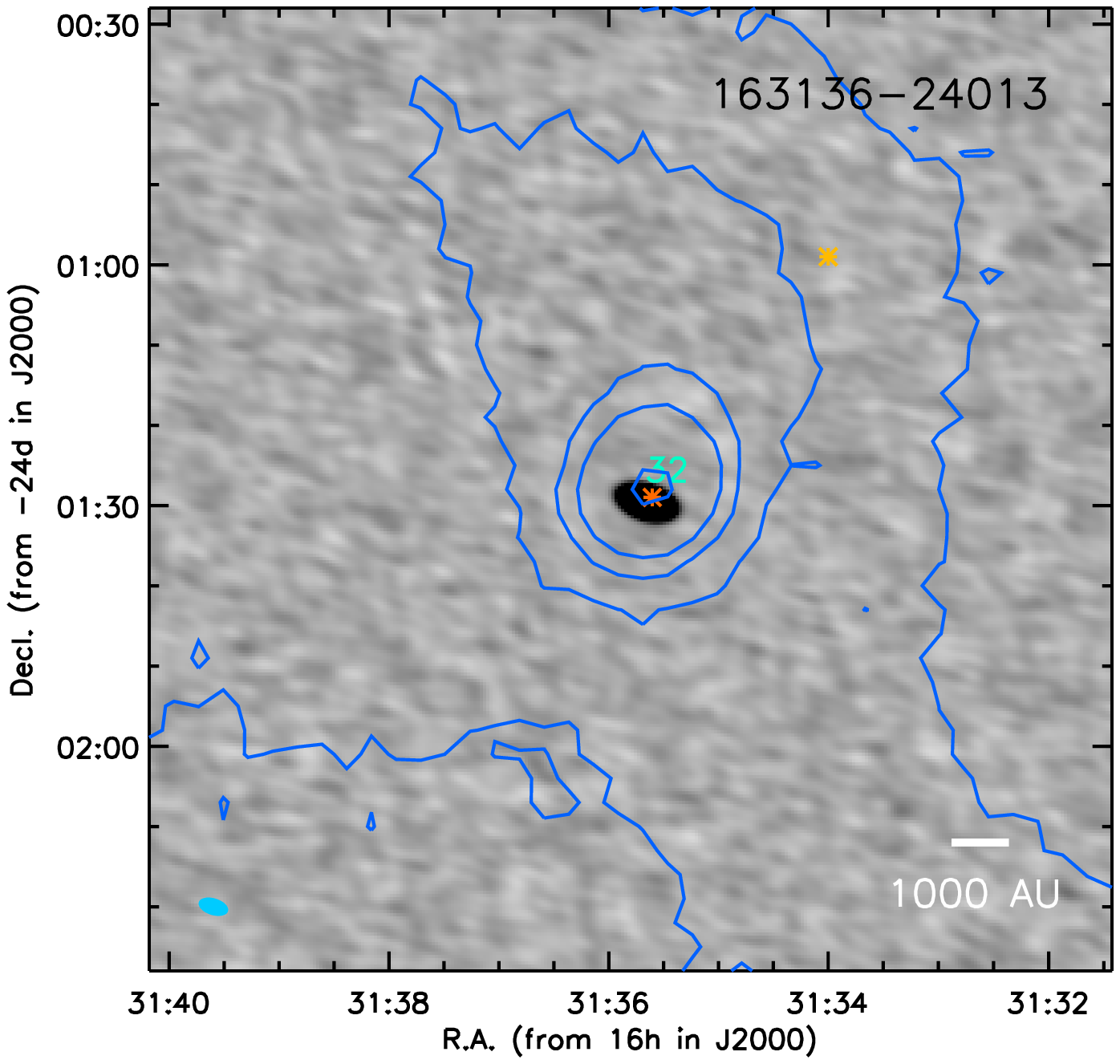} \\
\end{tabular}
\caption{ALMA detections coincident with a {\it Spitzer} YSO for six single-pointing
	fields.
	See Figures~\ref{fig_single_detects} 
	and \ref{fig_mosaic_detects} for the
	plotting conventions used.  
	}
\label{fig_proto_single1}
\end{figure}

\begin{figure}[p]
\begin{tabular}{cc}
\includegraphics[width=3in]{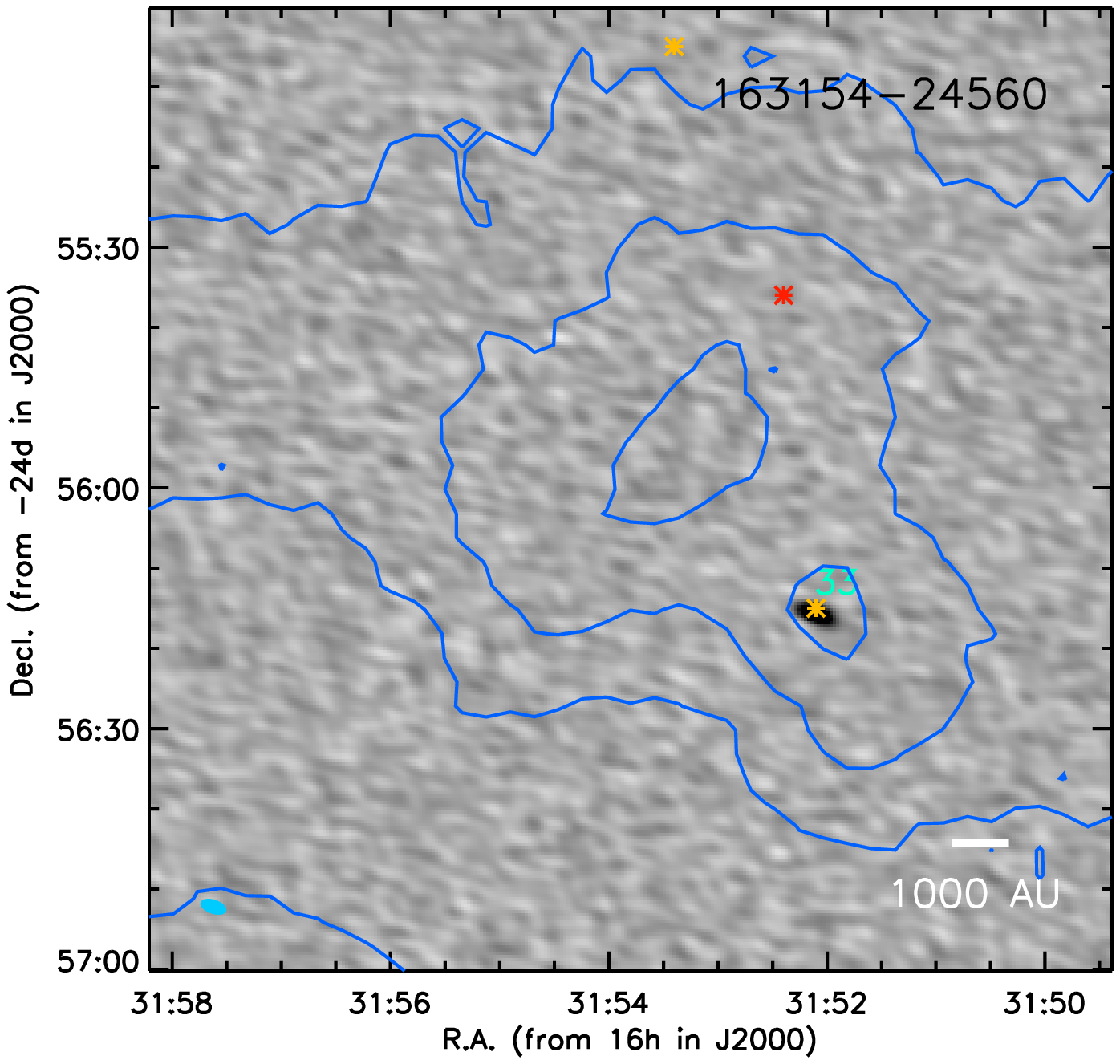} &
\includegraphics[width=3in]{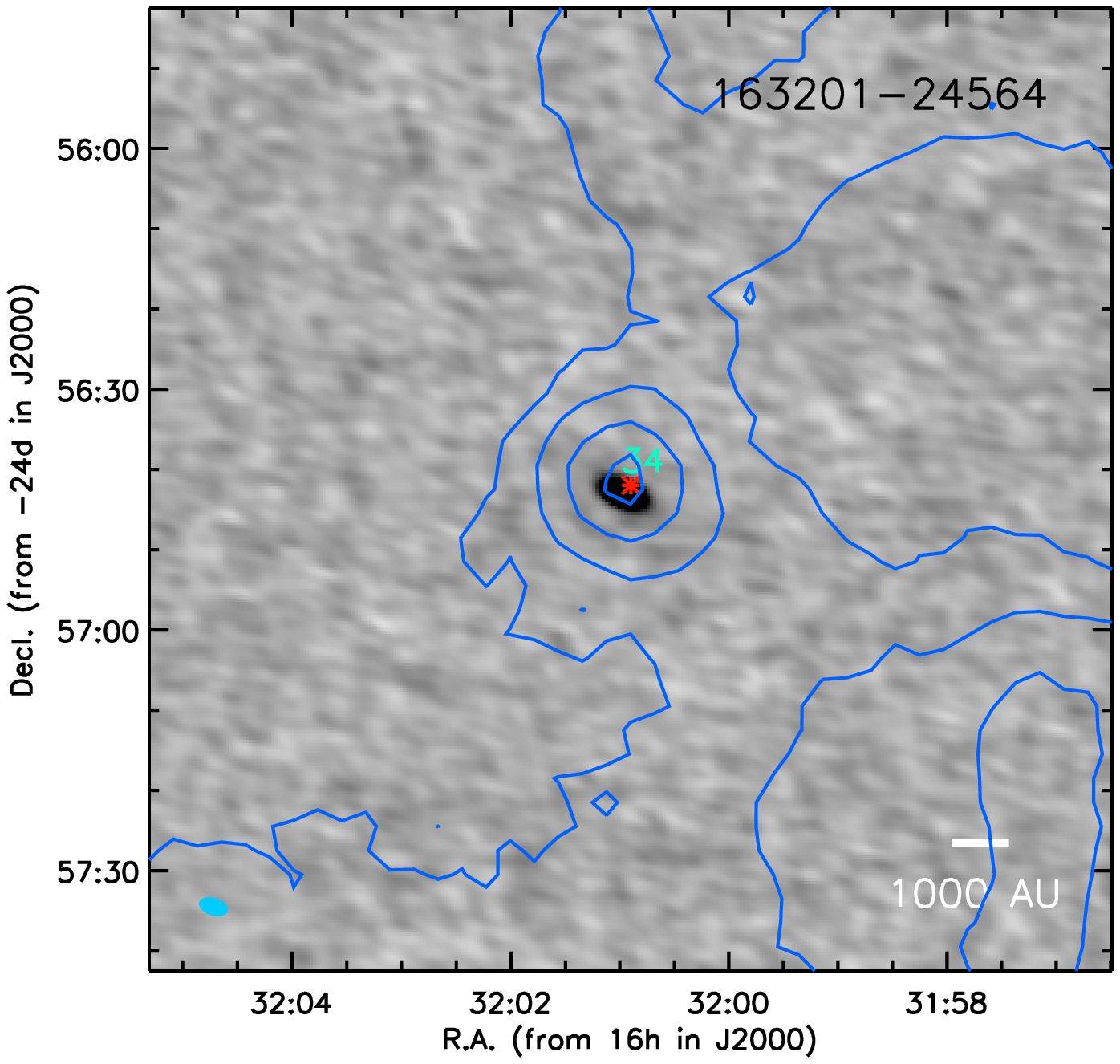} \\
\includegraphics[width=3in]{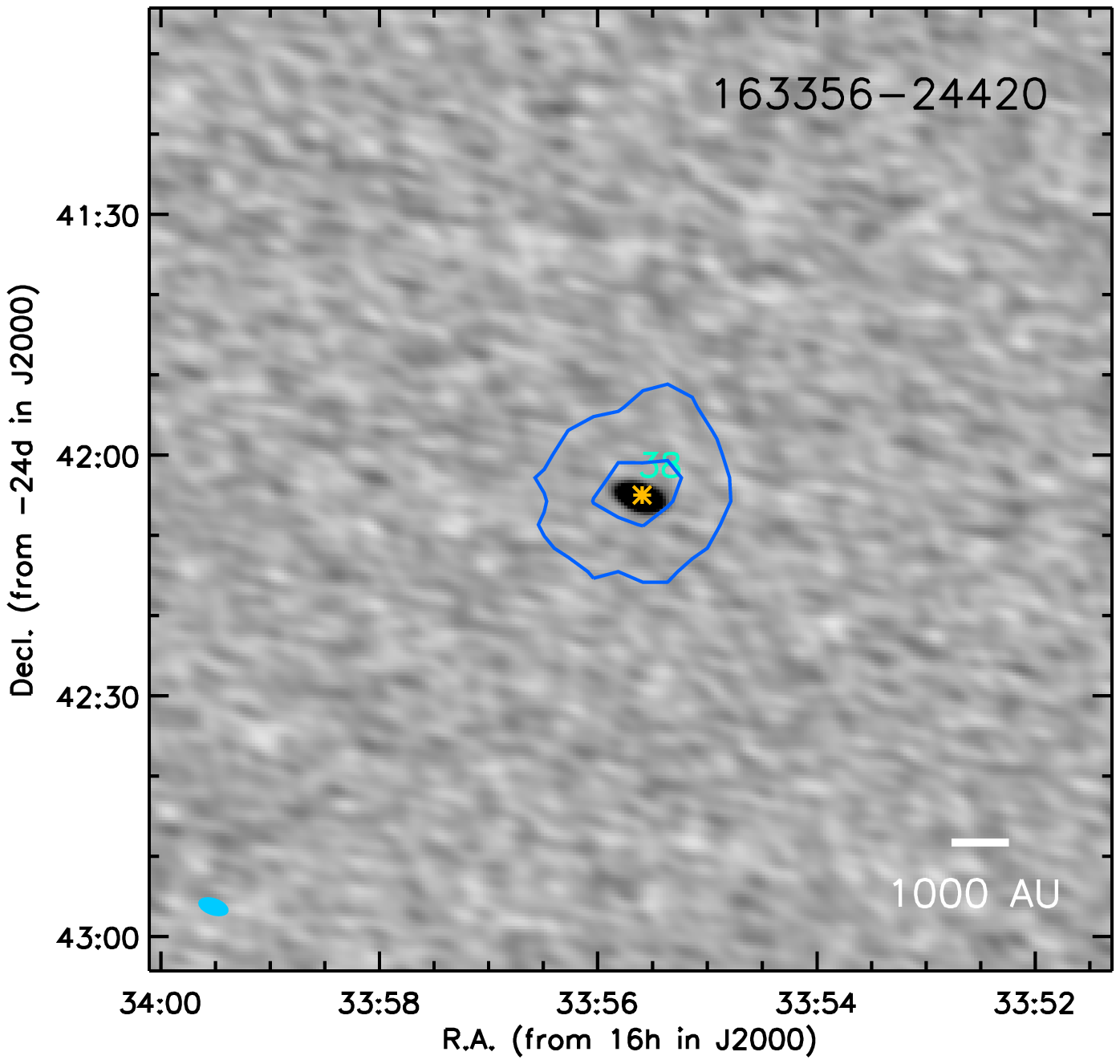} &
\\
\end{tabular}
\caption{ ALMA detections coincident with a {\it Spitzer} YSO for the remaining
	three single-pointing fields.
	See Figures~\ref{fig_single_detects} and \ref{fig_mosaic_detects}
	for the plotting conventions used.
	}
\label{fig_proto_single2}
\end{figure}

\begin{figure}[htb]
\includegraphics[width=4in]{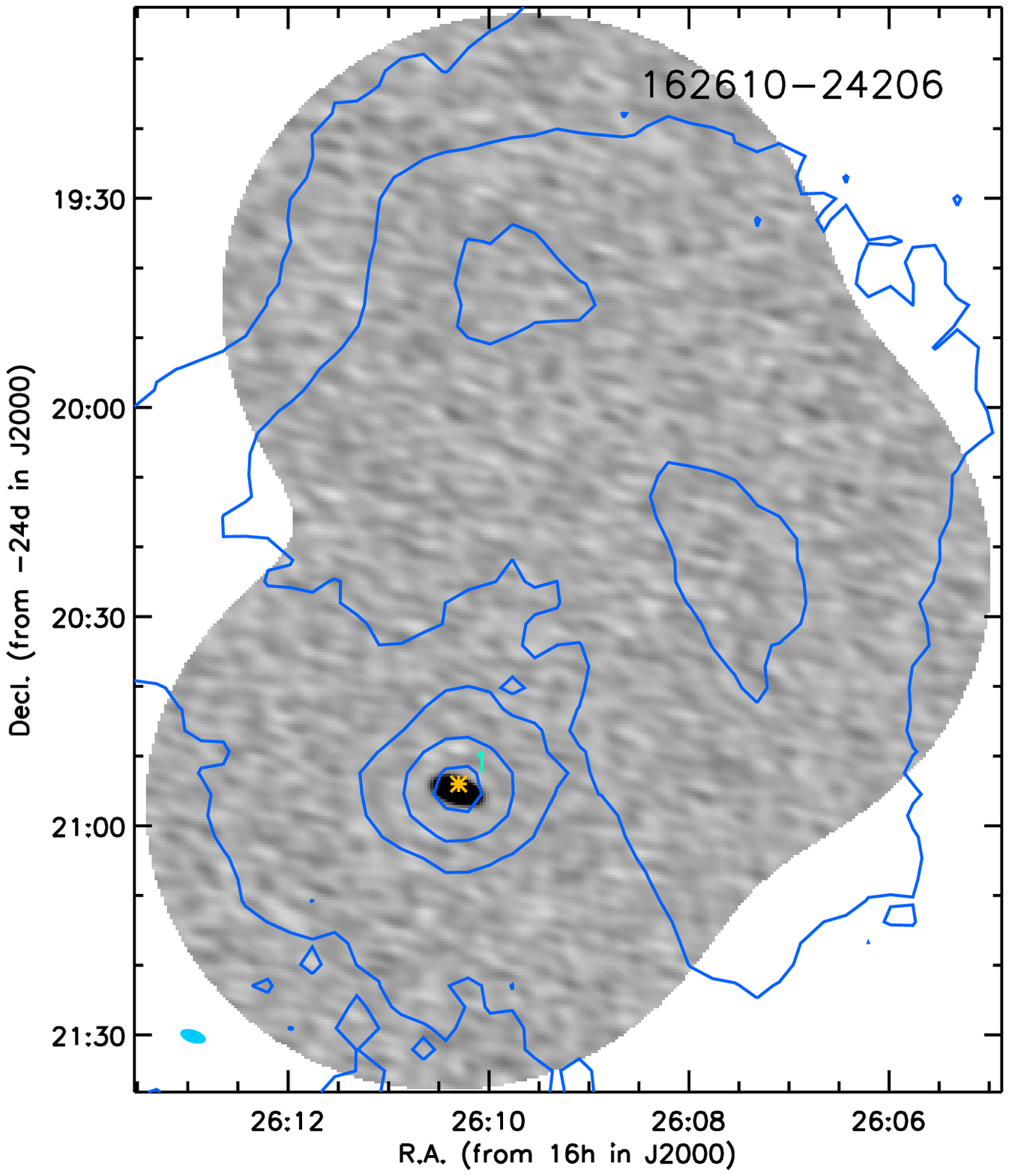}
\caption{ALMA detection coincident with a {\it Spitzer} YSO in the
	mosaicked field 162610-24206.
	See Figures~\ref{fig_single_detects} and \ref{fig_mosaic_detects} 
	for the plotting conventions used.} 
	
\label{fig_proto_mos1}
\end{figure}

\begin{figure}[htb!]
\includegraphics[width=4.0in]{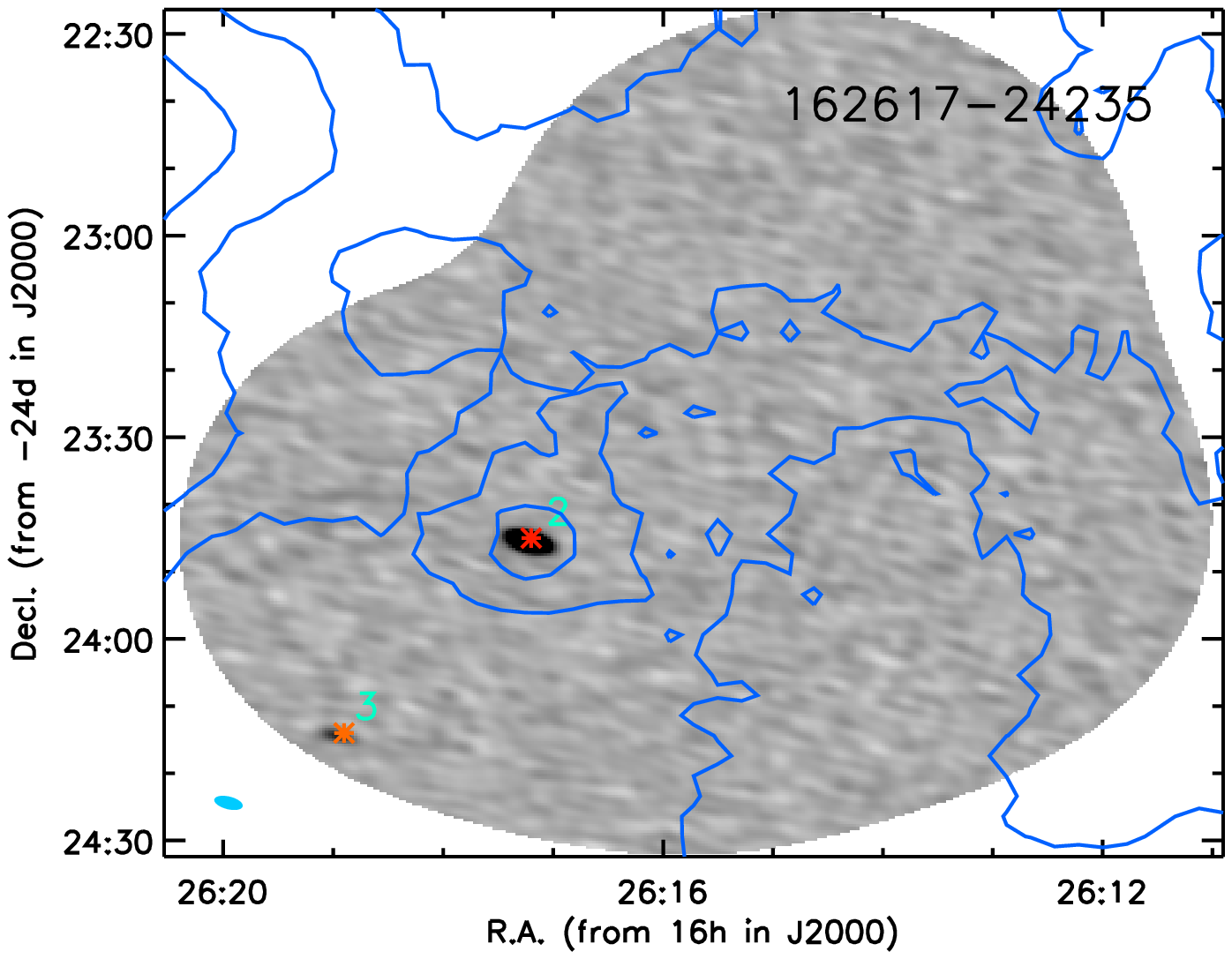}
\caption{ALMA detections coincident with {\it Spitzer} YSOs in the mosaicked
	field 162617-24235.
	See
	Figures~\ref{fig_single_detects} and \ref{fig_mosaic_detects} 
	for the plotting conventions used.}
\label{fig_proto_mos2}
\end{figure}

\begin{figure}[htb!]
\includegraphics[width=4in]{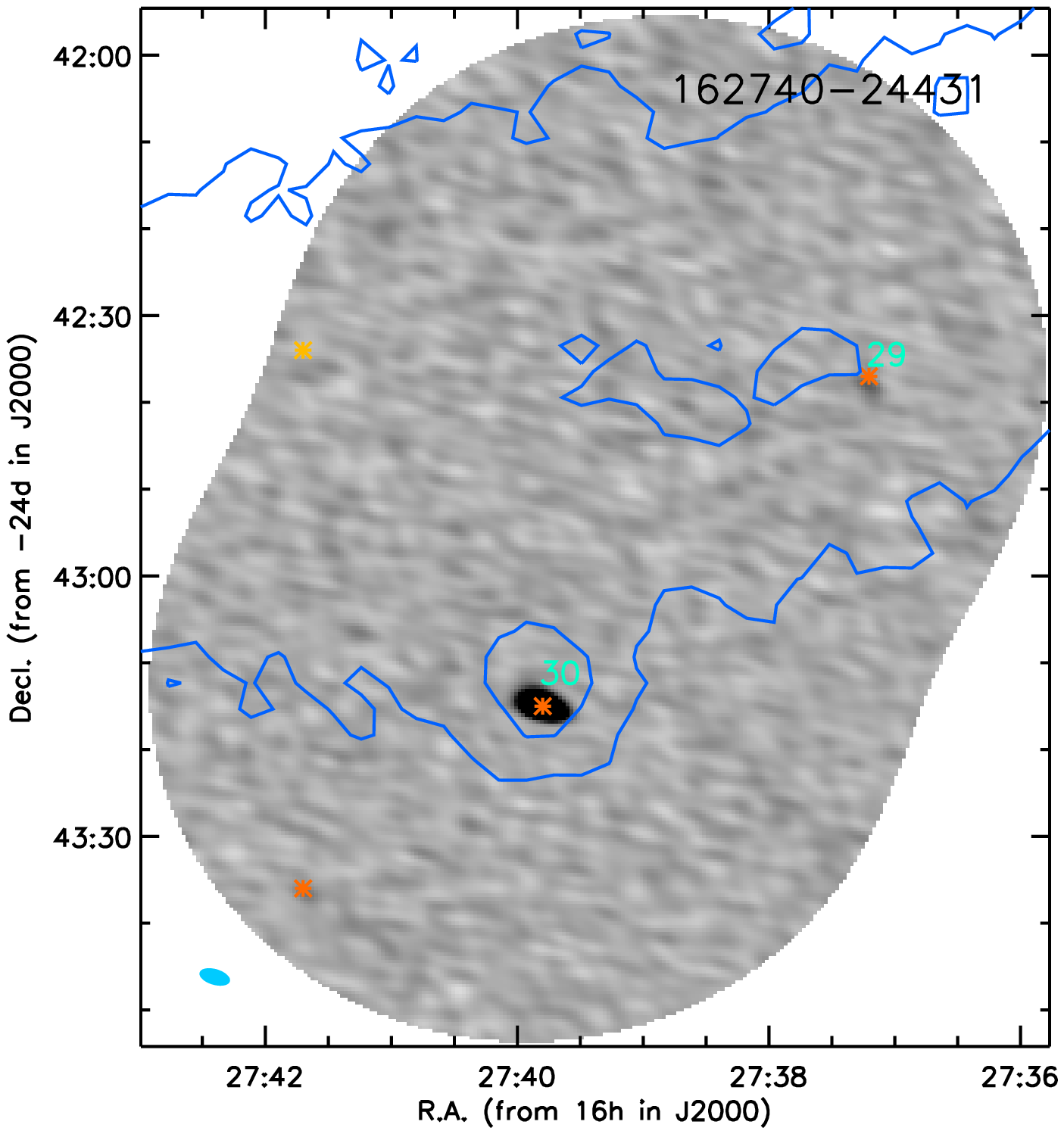}
\caption{ALMA detections coincident with {\it Spitzer} YSOs in the mosaicked
	field 162740-24431.
	See Figures~\ref{fig_single_detects} and \ref{fig_mosaic_detects} 
	for the plotting conventions used.}
\label{fig_proto_mos3}
\end{figure}

We have a large number of ALMA detections which are strongly associated 
(separations of $<2\arcsec$) with a {\it Spitzer} YSO.
Figure~\ref{fig_proto_single1} and \ref{fig_proto_single2} show single pointing 
fields, while Figures~\ref{fig_proto_mos1} to \ref{fig_proto_mos3} show
mosaicked areas.

As noted in Section~2, our initial sample contained 14, 4, and 5 SCUBA cores
associated with Class 0+I, Flat, and Class II YSOs, respectively.  Of these, we
detected 24, 3, and 5 sources with ALMA across 13, 3, and 5 unique cores, respectively.
In other words, only one Class 0+I and one Flat core had no ALMA detections within
the field of view.
Since earlier classes of protostars generally have more surrounding dust than
later classes, these two non-detections are surprising.  One possibility is that
these protostars are at a later stage of evolution, but are viewed edge-on, making
them appear younger.
As discussed earlier, some of the ALMA detections in the protostellar core 
fields are located a
significant distance from the nearest {\it Spitzer} YSO location.
Nearly every protostellar core, however, had at least 
one detection which was 
clearly associated with a
{\it Spitzer} YSO.  The lone exception is the Class~II core 162659-24454,
where ALMA detected a single emission peak located $\sim6$\arcsec\ from the nearest
{\it Spitzer} YSO listed in \citet{Dunham15}.  
This source is shown in Figure~\ref{fig_semioffs_protos1}.  Examination
of {\it Herschel} observations in this region suggest that the {\it Spitzer} catalogue
may have a positional offset, as {\it Herschel} emission 
from a YSO is centred on the ALMA detection.
We defer an in-depth examination of the properties of ALMA's protostellar
detections to a future paper.

\subsection{Protostellar Multiplicity}
The ALMA detections 
contains a number of pairs with projected separations of a few thousand au. These 
include previously identified wide protostellar binaries IRAS 16293-2422 
(sources 35 and 36 in Figure \ref{fig_semioffs_protos1}) and 
VLA 1623 and VLA 1623W (sources 8 and 9 in Figure \ref{fig_mosaic_detects}). 
Altogether 14 of our 38 
detections, or one third, have a second source with projected 
distance $\lesssim 3,000$ au. Figure \ref{fig_mosaic_detects} contains 
a particularly interesting example which could represent a forming
triple or a forming binary with an unbound single protostar.
Extended filamentary emission appears to connect all sources in the area,
also previously observed at larger angular scales by \citet{Motte98} and 
\citet{Johnstone00b}, for example,
which increases the confidence that they are forming together within a single structure.

Pairs with separations of $\sim$1,000-3,000~au may indicate the occurrence of turbulent 
fragmentation, which some 
numerical simulations predict may produce wide protobinaries on these 
size scales  \citep{Offner10,Offner16}. Observations of a wide-separation forming bound 
quadruple system was recently presented by \citet{Pineda15}, where more detailed velocity 
information was available. In a VLA survey of Perseus, \citet{Tobin16a} found more than 
half of identified protostellar pairs had projected separations greater than a few hundred au. 
They estimated the overall protostellar multiplicity fraction in Perseus is $0.40 \pm 0.06$. 
Thus, the fractions of potential wide-binaries in our Ophiuchus sample 
and Perseus are roughly consistent\footnote{Our ALMA observations are sensitive
to multiple systems with separations greater than several hundred au, and show a similar
distribution in the thousands of au range to the \citet{Tobin16a} Perseus results.}.

We caution, however, that projected proximity does not guarantee sources are gravitationally 
bound or will go on to form a binary star system.  Pairs may have more significant separations 
along the line-of-sight or be chance alignments. 

In addition, significant dynamical evolution, which impacts the separation distribution, is 
expected to occur during the protostellar phase \citep[e.g,][]{Bate02a,Tobin16a}. 
Sources with initially 
wide separations may migrate to shorter separations on $\sim0.1$Myr timescales or become 
unbound  \citep{Offner10,Offner16}. Velocity information is required to draw conclusions 
about the boundedness of close pairs in our sample and to determine whether such pairs 
are likely to remain bound \citep[e.g.,][]{Pineda15}.

\acknowledgements{
The authors thank the referee for their careful and constructive comments 
which improved our paper.
The National Radio Astronomy Observatory is a facility of the National Science Foundation 
operated under cooperative agreement by Associated Universities, Inc. This paper makes use 
of the following ALMA data: ADS/JAO.ALMA\# 2013.1.00187.S. ALMA is a partnership of ESO 
(representing its member states), NSF (USA), and NINS (Japan), together with NRC (Canada) 
and NSC and ASIAA (Taiwan), in cooperation with the Republic of Chile. The Joint ALMA 
Observatory is operated by ESO, AUI/NRAO, and NAOJ.
This research also made use of NASA's Astrophysics Data System (ADS) Abstract Service and 
the IDL Astronomy Library hosted by the NASA Goddard Space Flight Center.
This research has made use of the SIMBAD database, operated at CDS, Strasbourg, France
\citep{Wenger00}.  The JCMT has historically been operated by the Joint Astronomy Centre 
on behalf of the Science and Technology Facilities Council of the United Kingdom, the 
National Research Council of Canada and the Netherlands Organisation for Scientific Research. 
Additional funds for the construction of SCUBA-2 were provided by the Canada Foundation for 
Innovation. The identification number for the programme under which the SCUBA-2 data used 
in this paper is MJLSG32.
}

\facility{ALMA, {\it Spitzer}, JCMT, {\it Herschel}}
\software{CASA}

\bibliographystyle{apj}
\bibliography{bibfile}{}

\begin{thebibliography}{67}
\expandafter\ifx\csname natexlab\endcsname\relax\def\natexlab#1{#1}\fi

\bibitem[{{Andr{\'e}} {et~al.}(2010){Andr{\'e}}, {Men'shchikov}, {Bontemps},
  {K{\"o}nyves}, {Motte}, {Schneider}, {Didelon}, {Minier}, {Saraceno},
  {Ward-Thompson}, {di Francesco}, {White}, {Molinari}, {Testi}, {Abergel},
  {Griffin}, {Henning}, {Royer}, {Mer{\'{\i}}n}, {Vavrek}, {Attard},
  {Arzoumanian}, {Wilson}, {Ade}, {Aussel}, {Baluteau}, {Benedettini},
  {Bernard}, {Blommaert}, {Cambr{\'e}sy}, {Cox}, {di Giorgio}, {Hargrave},
  {Hennemann}, {Huang}, {Kirk}, {Krause}, {Launhardt}, {Leeks}, {Le Pennec},
  {Li}, {Martin}, {Maury}, {Olofsson}, {Omont}, {Peretto}, {Pezzuto}, {Prusti},
  {Roussel}, {Russeil}, {Sauvage}, {Sibthorpe}, {Sicilia-Aguilar}, {Spinoglio},
  {Waelkens}, {Woodcraft}, \& {Zavagno}}]{Andre10}
{Andr{\'e}}, P., {Men'shchikov}, A., {Bontemps}, S., {et~al.} 2010, \aap, 518,
  L102

\bibitem[{{Bate} {et~al.}(2002){Bate}, {Bonnell}, \& {Bromm}}]{Bate02a}
{Bate}, M.~R., {Bonnell}, I.~A., \& {Bromm}, V. 2002, \mnras, 336, 705

\bibitem[{{Belloche} {et~al.}(2011){Belloche}, {Schuller}, {Parise},
  {Andr{\'e}}, {Hatchell}, {J{\o}rgensen}, {Bontemps}, {Wei{\ss}}, {Menten}, \&
  {Muders}}]{Belloche11}
{Belloche}, A., {Schuller}, F., {Parise}, B., {et~al.} 2011, \aap, 527, A145

\bibitem[{{Bonnor}(1956)}]{Bonnor56}
{Bonnor}, W.~B. 1956, \mnras, 116, 351

\bibitem[{{Bourke} {et~al.}(2012){Bourke}, {Myers}, {Caselli}, {Di Francesco},
  {Belloche}, {Plume}, \& {Wilner}}]{Bourke12}
{Bourke}, T.~L., {Myers}, P.~C., {Caselli}, P., {et~al.} 2012, \apj, 745, 117

\bibitem[{{Chen} \& {Arce}(2010)}]{ChenArce10}
{Chen}, X., \& {Arce}, H.~G. 2010, \apjl, 720, L169

\bibitem[{{Chen} {et~al.}(2013){Chen}, {Arce}, {Zhang}, {Bourke}, {Launhardt},
  {J{\o}rgensen}, {Lee}, {Foster}, {Dunham}, {Pineda}, \& {Henning}}]{Chen13}
{Chen}, X., {Arce}, H.~G., {Zhang}, Q., {et~al.} 2013, \apj, 768, 110

\bibitem[{{Connelley} {et~al.}(2008){Connelley}, {Reipurth}, \&
  {Tokunaga}}]{Connelley08b}
{Connelley}, M.~S., {Reipurth}, B., \& {Tokunaga}, A.~T. 2008, \aj, 135, 2526

\bibitem[{{Di Francesco} {et~al.}(2007){Di Francesco}, {Evans}, {Caselli},
  {Myers}, {Shirley}, {Aikawa}, \& {Tafalla}}]{DiFrancesco07}
{Di Francesco}, J., {Evans}, II, N.~J., {Caselli}, P., {et~al.} 2007,
  Protostars and Planets V, 17

\bibitem[{{Duch{\^e}ne} \& {Kraus}(2013)}]{Duchene13}
{Duch{\^e}ne}, G., \& {Kraus}, A. 2013, \araa, 51, 269

\bibitem[{{Dunham} {et~al.}(2008){Dunham}, {Crapsi}, {Evans}, {Bourke},
  {Huard}, {Myers}, \& {Kauffmann}}]{Dunham08}
{Dunham}, M.~M., {Crapsi}, A., {Evans}, II, N.~J., {et~al.} 2008, \apjs, 179,
  249

\bibitem[{{Dunham} {et~al.}(2015){Dunham}, {Allen}, {Evans},
  {Broekhoven-Fiene}, {Cieza}, {Di Francesco}, {Gutermuth}, {Harvey},
  {Hatchell}, {Heiderman}, {Huard}, {Johnstone}, {Kirk}, {Matthews}, {Miller},
  {Peterson}, \& {Young}}]{Dunham15}
{Dunham}, M.~M., {Allen}, L.~E., {Evans}, II, N.~J., {et~al.} 2015, \apjs, 220,
  11

\bibitem[{{Dunham} {et~al.}(2016){Dunham}, {Offner}, {Pineda}, {Bourke},
  {Tobin}, {Arce}, {Chen}, {Di Francesco}, {Johnstone}, {Lee}, {Myers},
  {Price}, {Sadavoy}, \& {Schnee}}]{Dunham16}
{Dunham}, M.~M., {Offner}, S.~S.~R., {Pineda}, J.~E., {et~al.} 2016, \apj, 823,
  160

\bibitem[{{Duquennoy} \& {Mayor}(1991)}]{DuquennoyMayor91}
{Duquennoy}, A., \& {Mayor}, M. 1991, \aap, 248, 485

\bibitem[{{Ebert}(1955)}]{Ebert55}
{Ebert}, R. 1955, \zap, 37, 217

\bibitem[{{Enoch} {et~al.}(2008){Enoch}, {Evans}, {Sargent}, {Glenn},
  {Rosolowsky}, \& {Myers}}]{Enoch08}
{Enoch}, M.~L., {Evans}, II, N.~J., {Sargent}, A.~I., {et~al.} 2008, \apj, 684,
  1240

\bibitem[{{Evans} {et~al.}(2009){Evans}, {Dunham}, {J{\o}rgensen}, {Enoch},
  {Mer{\'{\i}}n}, {van Dishoeck}, {Alcal{\'a}}, {Myers}, {Stapelfeldt},
  {Huard}, {Allen}, {Harvey}, {van Kempen}, {Blake}, {Koerner}, {Mundy},
  {Padgett}, \& {Sargent}}]{Evans09}
{Evans}, II, N.~J., {Dunham}, M.~M., {J{\o}rgensen}, J.~K., {et~al.} 2009,
  \apjs, 181, 321

\bibitem[{{Fisher}(2004)}]{Fisher04}
{Fisher}, R.~T. 2004, \apj, 600, 769

\bibitem[{{Friesen} {et~al.}(2014){Friesen}, {Di Francesco}, {Bourke},
  {Caselli}, {J{\o}rgensen}, {Pineda}, \& {Wong}}]{Friesen14}
{Friesen}, R.~K., {Di Francesco}, J., {Bourke}, T.~L., {et~al.} 2014, \apj,
  797, 27

\bibitem[{{Friesen} {et~al.}(2009){Friesen}, {Di Francesco}, {Shirley}, \&
  {Myers}}]{Friesen09}
{Friesen}, R.~K., {Di Francesco}, J., {Shirley}, Y.~L., \& {Myers}, P.~C. 2009,
  \apj, 697, 1457

\bibitem[{{Gagn{\'e}} {et~al.}(2004){Gagn{\'e}}, {Skinner}, \&
  {Daniel}}]{Gagne04}
{Gagn{\'e}}, M., {Skinner}, S.~L., \& {Daniel}, K.~J. 2004, \apj, 613, 393

\bibitem[{{Goodwin} {et~al.}(2008){Goodwin}, {Nutter}, {Kroupa},
  {Ward-Thompson}, \& {Whitworth}}]{Goodwin08}
{Goodwin}, S.~P., {Nutter}, D., {Kroupa}, P., {Ward-Thompson}, D., \&
  {Whitworth}, A.~P. 2008, \aap, 477, 823

\bibitem[{{Goodwin} {et~al.}(2004){Goodwin}, {Whitworth}, \&
  {Ward-Thompson}}]{Goodwin04}
{Goodwin}, S.~P., {Whitworth}, A.~P., \& {Ward-Thompson}, D. 2004, \aap, 423,
  169

\bibitem[{{Holman} {et~al.}(2013){Holman}, {Walch}, {Goodwin}, \&
  {Whitworth}}]{Holman13}
{Holman}, K., {Walch}, S.~K., {Goodwin}, S.~P., \& {Whitworth}, A.~P. 2013,
  \mnras, 432, 3534

\bibitem[{{Jessop} \& {Ward-Thompson}(2000)}]{Jessop00}
{Jessop}, N.~E., \& {Ward-Thompson}, D. 2000, \mnras, 311, 63

\bibitem[{{Johnstone} {et~al.}(2004){Johnstone}, {Di Francesco}, \&
  {Kirk}}]{Johnstone04}
{Johnstone}, D., {Di Francesco}, J., \& {Kirk}, H. 2004, \apjl, 611, L45

\bibitem[{{Johnstone} {et~al.}(2000){Johnstone}, {Wilson}, {Moriarty-Schieven},
  {Joncas}, {Smith}, {Gregersen}, \& {Fich}}]{Johnstone00b}
{Johnstone}, D., {Wilson}, C.~D., {Moriarty-Schieven}, G., {et~al.} 2000, \apj,
  545, 327

\bibitem[{{J{\o}rgensen} {et~al.}(2007){J{\o}rgensen}, {Johnstone}, {Kirk}, \&
  {Myers}}]{Jorgensen07}
{J{\o}rgensen}, J.~K., {Johnstone}, D., {Kirk}, H., \& {Myers}, P.~C. 2007,
  \apj, 656, 293

\bibitem[{{J{\o}rgensen} {et~al.}(2008){J{\o}rgensen}, {Johnstone}, {Kirk},
  {Myers}, {Allen}, \& {Shirley}}]{Jorgensen08}
{J{\o}rgensen}, J.~K., {Johnstone}, D., {Kirk}, H., {et~al.} 2008, \apj, 683,
  822

\bibitem[{{Kainulainen} {et~al.}(2016){Kainulainen}, {Stutz}, {Stanke},
  {Abreu-Vicente}, {Beuther}, {Henning}, {Johnston}, \&
  {Megeath}}]{Kainulainen16}
{Kainulainen}, J., {Stutz}, A.~M., {Stanke}, T., {et~al.} 2016, ArXiv e-prints

\bibitem[{{Kirk} {et~al.}(2009){Kirk}, {Crutcher}, \&
  {Ward-Thompson}}]{JKirk09}
{Kirk}, J.~M., {Crutcher}, R.~M., \& {Ward-Thompson}, D. 2009, \apj, 701, 1044

\bibitem[{{Kratter} \& {Lodato}(2016)}]{Kratter16}
{Kratter}, K., \& {Lodato}, G. 2016, \araa, 54, 271

\bibitem[{{Lee} {et~al.}(2013){Lee}, {Looney}, {Schnee}, \& {Li}}]{Lee13}
{Lee}, K., {Looney}, L.~W., {Schnee}, S., \& {Li}, Z.-Y. 2013, \apj, 772, 100

\bibitem[{{Li} {et~al.}(2012){Li}, {Martin}, {Klein}, \& {McKee}}]{Li12}
{Li}, P.~S., {Martin}, D.~F., {Klein}, R.~I., \& {McKee}, C.~F. 2012, \apj,
  745, 139

\bibitem[{{Lin} {et~al.}(2012){Lin}, {Webb}, \& {Barret}}]{Lin12}
{Lin}, D., {Webb}, N.~A., \& {Barret}, D. 2012, \apj, 756, 27

\bibitem[{{Lis} {et~al.}(2016){Lis}, {Wootten}, {Gerin}, {Pagani}, {Roueff},
  {van der Tak}, {Vastel}, \& {Walmsley}}]{Lis16}
{Lis}, D.~C., {Wootten}, H.~A., {Gerin}, M., {et~al.} 2016, \apj, 827, 133

\bibitem[{{Lomax} {et~al.}(2015){Lomax}, {Whitworth}, {Hubber}, {Stamatellos},
  \& {Walch}}]{Lomax15}
{Lomax}, O., {Whitworth}, A.~P., {Hubber}, D.~A., {Stamatellos}, D., \&
  {Walch}, S. 2015, \mnras, 447, 1550

\bibitem[{{Looney} {et~al.}(2000){Looney}, {Mundy}, \& {Welch}}]{Looney00}
{Looney}, L.~W., {Mundy}, L.~G., \& {Welch}, W.~J. 2000, \apj, 529, 477

\bibitem[{{Mairs} {et~al.}(2014){Mairs}, {Johnstone}, {Offner}, \&
  {Schnee}}]{Mairs14}
{Mairs}, S., {Johnstone}, D., {Offner}, S.~S.~R., \& {Schnee}, S. 2014, \apj,
  783, 60

\bibitem[{{Molinari} {et~al.}(2011){Molinari}, {Schisano}, {Faustini},
  {Pestalozzi}, {di Giorgio}, \& {Liu}}]{Molinari11}
{Molinari}, S., {Schisano}, E., {Faustini}, F., {et~al.} 2011, \aap, 530, A133

\bibitem[{{Motte} {et~al.}(1998){Motte}, {Andre}, \& {Neri}}]{Motte98}
{Motte}, F., {Andre}, P., \& {Neri}, R. 1998, \aap, 336, 150

\bibitem[{{Nakamura} {et~al.}(2012){Nakamura}, {Takakuwa}, \&
  {Kawabe}}]{Nakamura12}
{Nakamura}, F., {Takakuwa}, S., \& {Kawabe}, R. 2012, \apjl, 758, L25

\bibitem[{{Nutter} {et~al.}(2006){Nutter}, {Ward-Thompson}, \&
  {Andr{\'e}}}]{Nutter06}
{Nutter}, D., {Ward-Thompson}, D., \& {Andr{\'e}}, P. 2006, \mnras, 368, 1833

\bibitem[{{Offner} {et~al.}(2012){Offner}, {Capodilupo}, {Schnee}, \&
  {Goodman}}]{Offner12}
{Offner}, S.~S.~R., {Capodilupo}, J., {Schnee}, S., \& {Goodman}, A.~A. 2012,
  \mnras, 420, L53

\bibitem[{{Offner} {et~al.}(2016){Offner}, {Dunham}, {Lee}, {Arce}, \&
  {Fielding}}]{Offner16}
{Offner}, S.~S.~R., {Dunham}, M.~M., {Lee}, K.~I., {Arce}, H.~G., \&
  {Fielding}, D.~B. 2016, \apjl, 827, L11

\bibitem[{{Offner} {et~al.}(2010){Offner}, {Kratter}, {Matzner}, {Krumholz}, \&
  {Klein}}]{Offner10}
{Offner}, S.~S.~R., {Kratter}, K.~M., {Matzner}, C.~D., {Krumholz}, M.~R., \&
  {Klein}, R.~I. 2010, \apj, 725, 1485

\bibitem[{{Ortiz-Le{\'o}n} {et~al.}(2016){Ortiz-Le{\'o}n}, {Loinard},
  {Kounkel}, {Dzib}, {Mioduszewski}, {Rodr{\'{\i}}guez}, {Torres},
  {Gonz{\'a}lez-L{\'o}pezlira}, {Pech}, {Rivera}, {Hartmann}, {Boden}, {Evans},
  {Brice{\~n}o}, {Tobin}, {Galli}, \& {Gudehus}}]{OrtizLeon17}
{Ortiz-Le{\'o}n}, G.~N., {Loinard}, L., {Kounkel}, M.~A., {et~al.} 2016, ArXiv
  e-prints

\bibitem[{{Ossenkopf} \& {Henning}(1994)}]{Ossenkopf94}
{Ossenkopf}, V., \& {Henning}, T. 1994, \aap, 291, 943

\bibitem[{{Padgett} {et~al.}(2008){Padgett}, {Rebull}, {Stapelfeldt},
  {Chapman}, {Lai}, {Mundy}, {Evans}, {Brooke}, {Cieza}, {Spiesman},
  {Noriega-Crespo}, {McCabe}, {Allen}, {Blake}, {Harvey}, {Huard},
  {J{\o}rgensen}, {Koerner}, {Myers}, {Sargent}, {Teuben}, {van Dishoeck},
  {Wahhaj}, \& {Young}}]{Padgett08}
{Padgett}, D.~L., {Rebull}, L.~M., {Stapelfeldt}, K.~R., {et~al.} 2008, \apj,
  672, 1013

\bibitem[{{Pattle} {et~al.}(2015){Pattle}, {Ward-Thompson}, {Kirk}, {White},
  {Drabek-Maunder}, {Buckle}, {Beaulieu}, {Berry}, {Broekhoven-Fiene},
  {Currie}, {Fich}, {Hatchell}, {Kirk}, {Jenness}, {Johnstone}, {Mottram},
  {Nutter}, {Pineda}, {Quinn}, {Salji}, {Tisi}, {Walker-Smith}, {Francesco},
  {Hogerheijde}, {Andr{\'e}}, {Bastien}, {Bresnahan}, {Butner}, {Chen},
  {Chrysostomou}, {Coude}, {Davis}, {Duarte-Cabral}, {Fiege}, {Friberg},
  {Friesen}, {Fuller}, {Graves}, {Greaves}, {Gregson}, {Griffin}, {Holland},
  {Joncas}, {Knee}, {K{\"o}nyves}, {Mairs}, {Marsh}, {Matthews},
  {Moriarty-Schieven}, {Rawlings}, {Richer}, {Robertson}, {Rosolowsky},
  {Rumble}, {Sadavoy}, {Spinoglio}, {Thomas}, {Tothill}, {Viti}, {Wouterloot},
  {Yates}, \& {Zhu}}]{Pattle15}
{Pattle}, K., {Ward-Thompson}, D., {Kirk}, J.~M., {et~al.} 2015, \mnras, 450,
  1094

\bibitem[{{Pineda} {et~al.}(2011){Pineda}, {Goodman}, {Arce}, {Caselli},
  {Longmore}, \& {Corder}}]{Pineda11b}
{Pineda}, J.~E., {Goodman}, A.~A., {Arce}, H.~G., {et~al.} 2011, \apjl, 739, L2

\bibitem[{{Pineda} {et~al.}(2009){Pineda}, {Rosolowsky}, \&
  {Goodman}}]{Pineda09}
{Pineda}, J.~E., {Rosolowsky}, E.~W., \& {Goodman}, A.~A. 2009, \apjl, 699,
  L134

\bibitem[{{Pineda} {et~al.}(2015){Pineda}, {Offner}, {Parker}, {Arce},
  {Goodman}, {Caselli}, {Fuller}, {Bourke}, \& {Corder}}]{Pineda15}
{Pineda}, J.~E., {Offner}, S.~S.~R., {Parker}, R.~J., {et~al.} 2015, \nat, 518,
  213

\bibitem[{{Raghavan} {et~al.}(2010){Raghavan}, {McAlister}, {Henry}, {Latham},
  {Marcy}, {Mason}, {Gies}, {White}, \& {ten Brummelaar}}]{Raghavan10}
{Raghavan}, D., {McAlister}, H.~A., {Henry}, T.~J., {et~al.} 2010, \apjs, 190,
  1

\bibitem[{{Sadavoy} {et~al.}(2010{\natexlab{a}}){Sadavoy}, {Di Francesco}, \&
  {Johnstone}}]{Sadavoy10b}
{Sadavoy}, S.~I., {Di Francesco}, J., \& {Johnstone}, D. 2010{\natexlab{a}},
  \apjl, 718, L32

\bibitem[{{Sadavoy} {et~al.}(2010{\natexlab{b}}){Sadavoy}, {Di Francesco},
  {Bontemps}, {Megeath}, {Rebull}, {Allgaier}, {Carey}, {Gutermuth}, {Hora},
  {Huard}, {McCabe}, {Muzerolle}, {Noriega-Crespo}, {Padgett}, \&
  {Terebey}}]{Sadavoy10a}
{Sadavoy}, S.~I., {Di Francesco}, J., {Bontemps}, S., {et~al.}
  2010{\natexlab{b}}, \apj, 710, 1247

\bibitem[{{Schnee} {et~al.}(2012){Schnee}, {Di Francesco}, {Enoch}, {Friesen},
  {Johnstone}, \& {Sadavoy}}]{Schnee12a}
{Schnee}, S., {Di Francesco}, J., {Enoch}, M., {et~al.} 2012, \apj, 745, 18

\bibitem[{{Schnee} {et~al.}(2010){Schnee}, {Enoch}, {Johnstone}, {Culverhouse},
  {Leitch}, {Marrone}, \& {Sargent}}]{Schnee10}
{Schnee}, S., {Enoch}, M., {Johnstone}, D., {et~al.} 2010, \apj, 718, 306

\bibitem[{{Shinnaga} {et~al.}(2015){Shinnaga}, {Humphreys}, {Indebetouw},
  {Villard}, {Kern}, {Davis}, {Miura}, {Nakazato}, {Sugimoto}, {Kosugi},
  {Akiyama}, {Muders}, {Wyrowski}, {Williams}, {Lightfoot}, {Kent}, {Momjian},
  {Hunter}, \& {ALMA Pipeline Team}}]{Shinnaga15}
{Shinnaga}, H., {Humphreys}, E., {Indebetouw}, R., {et~al.} 2015, in
  Astronomical Society of the Pacific Conference Series, Vol. 499, Revolution
  in Astronomy with ALMA: The Third Year, ed. D.~{Iono}, K.~{Tatematsu},
  A.~{Wootten}, \& L.~{Testi}, 355

\bibitem[{{Stanke} {et~al.}(2006){Stanke}, {Smith}, {Gredel}, \&
  {Khanzadyan}}]{Stanke06}
{Stanke}, T., {Smith}, M.~D., {Gredel}, R., \& {Khanzadyan}, T. 2006, \aap,
  447, 609

\bibitem[{{Stutz} {et~al.}(2013){Stutz}, {Tobin}, {Stanke}, {Megeath},
  {Fischer}, {Robitaille}, {Henning}, {Ali}, {di Francesco}, {Furlan},
  {Hartmann}, {Osorio}, {Wilson}, {Allen}, {Krause}, \& {Manoj}}]{Stutz13}
{Stutz}, A.~M., {Tobin}, J.~J., {Stanke}, T., {et~al.} 2013, \apj, 767, 36

\bibitem[{{Tobin} {et~al.}(2016{\natexlab{a}}){Tobin}, {Kratter}, {Persson},
  {Looney}, {Dunham}, {Segura-Cox}, {Li}, {Chandler}, {Sadavoy}, {Harris},
  {Melis}, \& {P{\'e}rez}}]{Tobin16c}
{Tobin}, J.~J., {Kratter}, K.~M., {Persson}, M.~V., {et~al.}
  2016{\natexlab{a}}, \nat, 538, 483

\bibitem[{{Tobin} {et~al.}(2016{\natexlab{b}}){Tobin}, {Looney}, {Li},
  {Chandler}, {Dunham}, {Segura-Cox}, {Sadavoy}, {Melis}, {Harris}, {Kratter},
  \& {Perez}}]{Tobin16a}
{Tobin}, J.~J., {Looney}, L.~W., {Li}, Z.-Y., {et~al.} 2016{\natexlab{b}},
  \apj, 818, 73

\bibitem[{{Wenger} {et~al.}(2000){Wenger}, {Ochsenbein}, {Egret}, {Dubois},
  {Bonnarel}, {Borde}, {Genova}, {Jasniewicz}, {Lalo{\"e}}, {Lesteven}, \&
  {Monier}}]{Wenger00}
{Wenger}, M., {Ochsenbein}, F., {Egret}, D., {et~al.} 2000, \aaps, 143, 9

\bibitem[{{Williams} {et~al.}(1994){Williams}, {de Geus}, \&
  {Blitz}}]{Williams94}
{Williams}, J.~P., {de Geus}, E.~J., \& {Blitz}, L. 1994, \apj, 428, 693

\bibitem[{{Young} {et~al.}(2006){Young}, {Enoch}, {Evans}, {Glenn}, {Sargent},
  {Huard}, {Aguirre}, {Golwala}, {Haig}, {Harvey}, {Laurent}, {Mauskopf}, \&
  {Sayers}}]{Young06}
{Young}, K.~E., {Enoch}, M.~L., {Evans}, II, N.~J., {et~al.} 2006, \apj, 644,
  326

\bibitem[{{Zhang} {et~al.}(2013){Zhang}, {Brandner}, {Wang}, {Gennaro}, {Bik},
  {Henning}, {Gredel}, {Smith}, \& {Stanke}}]{Zhang13}
{Zhang}, M., {Brandner}, W., {Wang}, H., {et~al.} 2013, \aap, 553, A41

\end{thebibliography}

\end{document}